\documentclass{aa}  

\usepackage{graphicx}
\usepackage{txfonts}
\usepackage{lipsum}
\usepackage{subcaption}         % necessary for continued figures, example in section 3
\usepackage{lscape}             % to rotate a single page table, example in appendix.
\usepackage{placeins}           % useful with \FloatBarrier, to keep 
\usepackage{physics,natbib,tabularx,xcolor}
\usepackage{hyperref,multirow}
\hypersetup{% hyperref option list
setpagesize=false,
 bookmarksnumbered=true,%
 bookmarksopen=true,%
 colorlinks=true,%
linkcolor=blue,
citecolor=blue,
}
\newcommand{\rev}[1]{\textcolor{black}{#1}}

\begin{document}

   \title{\rev{Rapid and Predictive Planet Population Synthesis Model (RAPPS)}
   }
   \subtitle{I. Upgraded model and resulting synthetic populations}
    %%%%%%%%%%%%%%%%%%%%%%%%%%%%%%%%%%%%%%%%
    % Please do not include ORCIDs next to author names.
    % Only ORCIDs authenticated by individual authors in EDP Sciences editorial system will be taken into account.
    % ORCIDs included here will be removed.
    %%%%%%%%%%%%%%%%%%%%%%%%%%%%%%%%%%%%%%%%
   \author{Tadahiro Kimura
          \inst{1,2,3}
          \and
          Masahiro Ikoma\inst{3,4,5,6}
          % \fnmsep\thanks{Just to show the usage
          % of the elements in the author field}
          }

   \institute{
    Kapteyn Astronomical Institute, University of Groningen Landleven 12, 9747 AD, Groningen, Netherlands \\
    \email{tad.kimura624@gmail.com}    
    \and
    UTokyo Organization for Planetary Space Science (UTOPS), University of Tokyo, Hongo, Bunkyo-ku, Tokyo 113-0033, Japan
    \and
    Division of Science, National Astronomical Observatory of Japan, 
    Mitaka, Tokyo 181-8588, Japan   
    \and
    Astrobiology Center (ABC), Mitaka, Tokyo 181-8588, Japan   
    \and
    Graduate Institute for Advanced Studies, SOKENDAI, Mitaka, Tokyo 181-8588, Japan
    \and
    Department of Earth and Planetary Science, University of Tokyo, Hongo, Bunkyo-ku, Tokyo 113-0033, Japan
             }

   \date{Received September 15, 1996; accepted March 16, 1997}

% \abstract{}{}{}{}{} 
% 5 {} token are mandatory
 
  \abstract
    {Exoplanet surveys have revealed a large diversity of planetary systems. %, with planets spanning a wide range of masses, radii, and orbital configurations. 
    Explaining the origin of such diversity requires integrated models of planet formation. % that account for disc structure and evolution, core growth, atmospheric accumulation and escape, orbital migration, and dynamical evolution including resonance trapping and giant collisions. 
    Planet population synthesis (PPS) modelling is a key tool for linking theories with the statistical properties of observed exoplanets.
    In the upcoming decade, the number of known exoplanets is expected to increase ten-fold, and the range of planetary parameters will expand significantly due to near-future missions such as Roman, PLATO, and Ariel.
    }
    {We aim to develop a new PPS model capable of predicting planetary masses, radii, orbits, and atmospheric properties across a wide range of stellar hosts. 
    Given the expected significant increase in observed planetary data, %expected from upcoming missions, 
    the model must also achieve sufficiently high computational efficiency to allow a large number of simulations for statistical comparison with observational results.}
    {In this study, we build upon our previous PPS model \citep{Kimura+Ikoma2022}, which introduced the effects of water enrichment of primordial atmospheres through magma–atmosphere interactions, enhancing it to include a semi-analytical model for dynamical evolution of multiple-planet systems after disc dispersal; the model was demonstrated to reproduce the results of direct $N$-body simulations \citep{Kimura+etal2025a,Kimura+2025b}. 
    Additional updates include revised prescriptions for disc gas evolution, resonance trapping, and atmospheric escape.}
    {We show that our updated model produces planetary distributions that differ from our previous results, particularly in the abundance of Earth-mass and sub-Earth-mass planets. 
    These differences arise mainly from the new dynamical evolution treatment, and the resulting distributions are more consistent with other formation simulations using direct $N$-body integrations. 
    Our simulations also demonstrate that enrichment of primordial atmospheres through magma–atmosphere interaction strongly influences both the occurrence of gas giants and the radius distribution of close-in super-Earths and sub-Neptunes.}
    % conclusions heading (optional)
    {The upgraded PPS model provides a computationally efficient and physically comprehensive framework for predicting planetary properties across diverse stellar types. 
    It enables large parameter surveys and robust statistical comparisons with observational data, thereby offering the foundation for future detailed studies and for model validation against exoplanet observations.}


   \maketitle
   \nolinenumbers
%
%-------------------------------------------------------------------

%--------------------------------------------------------------------
\section{Introduction}
Over the past three decades, exoplanet surveys have revealed that planets are ubiquitous in the Milky Way Galaxy and exhibit remarkable diversity in their orbits, masses, and radii~\citep[e.g.,][]{Howard2013,Batalha2014,Zhu+2018}.
Although the first detection was a short-period gas giant~\citep{Mayor+Queloz1995}, transit and radial-velocity surveys have since established that planets with radii between those of Earth and Neptune are among the most common, especially at orbital periods shorter than $\sim$100~days \citep[see][for recent reviews]{Winn+Fabrycky2015,Zhu+Dong2021}. 
These findings demonstrate that planetary systems are markedly different from our Solar System, providing key constraints on theory of planet formation and evolution.

Planet formation theory has achieved substantial progress in identifying relevant physical processes, from dust coagulation in protoplanetary discs to core formation, atmospheric accumulation and escape, orbital migration, and late-stage dynamical evolution. 
Observational advances with facilities such as ALMA have also shed light on the initial conditions of planet formation by revealing disc structures \citep[e.g.,][]{Ansdell+2018,Williams+etal2019,Andrews2020,Tobin+2020}. 
However, many processes remain difficult to constrain directly, and testing their combined outcomes against the observed planetary populations remains a central challenge.

Planet population synthesis (PPS) offers a powerful framework for connecting planet formation theories with exoplanet observations. 
In this approach, various physical processes are combined into a single model that produces synthetic populations for direct comparison with observational results. 
Since the pioneering study done by \citet{Ida+Lin2004}, PPS models have been extensively developed and improved by several different groups \citep[e.g.,][among others]{Alibert+2005,Mordasini+2009,Mordasini+2012,Ida+2013,Coleman+Nelson2014,Bitsch+2015b,Ronco+2017,Brugger+2020,Emsenhuber+2021a}; see \cite{Burn+Mordasini2025} for a recent review.

Despite this progress, most comparisons between PPS predictions and observed planet distributions remain largely qualitative. 
Quantitative statistical studies have been limited \citep[e.g.,][]{Mordasini+2009b,Alibert+2011,Mishra+2021}. This is primarily due to computational costs and model limitations. 
With ongoing and upcoming missions such as TESS~\citep{Ricker+2015}, JWST~\citep{Gardner+2006}, PLATO~\citep{Rauer+2014}, Roman~\citep{Spergel+2015}, and Ariel~\citep{Tinetti+2016}, the observational scope will expand to include long-period planets, a wider variety of host stars, and atmospheric characterisation. 
Comparing theoretical predictions with such huge amounts of data will be essential to test and advance planet formation theory. 
To meet this challenge, next-generation PPS models must be capable of predicting diverse planetary observables---orbital periods, masses, radii, and atmospheric compositions---across a wide range of stellar types, while maintaining computational efficiency. 
%In addition, understanding processes such as water delivery is crucial for exploring the formation of potentially habitable planets. 

One of the most computationally expensive steps in numerical calculations for PPS is the late-stage dynamical evolution of multi-planet systems. In our previous work \citep[][hereafter KI22]{Kimura+Ikoma2022}, we developed a PPS model based on \cite{Ida+2013} and \cite{Emsenhuber+2021a}, aiming to predict the occurrence rates of potentially habitable planets with a particular focus on the effects of water enrichment in primordial atmospheres through magma–atmosphere interaction, which was proposed by \cite{Ikoma+Genda2006}. 
%We showed that this process strongly affects the final water inventory of terrestrial planets.
However, the semi-analytical treatment of planet–planet dynamical interactions adopted in that model tends to significantly underestimate the frequency of giant impacts after disc dispersal, particularly for close-in planets, compared to the results from direct $N$-body simulations. This limitation is critical, because late-stage giant impacts play an essential role in shaping the masses and orbital architectures of terrestrial planets and super-Earths.
%was not sufficiently accurate for, 
%even though these processes are critical for shaping terrestrial planets and super-Earths. 

Here we present an upgraded PPS model that builds on KI22. 
The primary improvement is the incorporation of a newly developed semi-analytical model for dynamical evolution of multi-planet systems after disc dispersal, which was developed by \citet{Kimura+etal2025a} and \citet{Kimura+2025b}; the semi-analytical model captures gravitational scattering and giant impacts without relying on direct $N$-body simulations. 
This enables a computationally efficient yet physically realistic treatment of dynamical evolution following disc dispersal. 
Additional updates include refined prescriptions for disc gas evolution, resonance trapping, and atmospheric escape. 
Collectively, these developments yield a PPS model capable of predicting planetary orbits, masses, radii, and atmospheric properties across a wide range of stellar hosts, while maintaining the computational speed necessary for large-scale statistical analyses.

The main purpose of this paper is to describe the details of our updated model and demonstrate its capabilities through representative simulations. 
We highlight how the improvements alter simulation outcomes relative to KI22, and examine how the resulting planetary distributions depend on stellar mass and primordial atmospheric composition. 
We also provide a comparison with the PPS model of \cite{Emsenhuber+2021a}. 
More comprehensive parameter studies and detailed statistical comparisons with observations will be presented in subsequent papers.

The remainder of this paper is organized as follows. 
In \S~\ref{sec:method}, we describe the components of our updated PPS model, focusing on the improvements relative to KI22. 
In \S~\ref{sec:example_simulation}, we present representative simulations to illustrate how planets and planetary systems form and evolve within our framework. 
In \S~\ref{sec:Monte_Carlo}, we show the results of population synthesis runs, examining how the updated model affects the final outcomes compared to KI22 and how the resulting distributions depend on stellar mass and primordial atmospheric composition. 
In \S~\ref{sec:discussion}, we compare our results with those of \cite{Emsenhuber+2021a} and also revisit the KI22 findings regarding terrestrial water content, as well as briefly discussing the caveats of our model. 
Finally, in \S~\ref{sec:conclusion}, we summarise the paper.

\section{Method}
\label{sec:method}

Here, we describe the details of our planet population synthesis (PPS) model, which is based on the planetesimal-driven core accretion scenario. 
The model builds on KI22, with major updates to the treatment of disc structure and evolution, and to dynamical interactions between planets.
The main physical processes incorporated in our model, along with the updates relative to KI22, are summarised below:
\begin{itemize}
    \item \textbf{Central star evolution} (\S~\ref{sec:star}) \\
    We adopt updated tabulated data for stellar bolometric evolution. The treatment of high-energy radiation (X-ray and EUV) remains unchanged from KI22.
    \item \textbf{Protoplanetary disc evolution} (\S~\ref{sec:disc}) \\
    The photoevaporation model has been significantly revised. Minor updates have also been made to the initial distributions of gas and solids and to the mid-plane temperature model.
    \item \textbf{Solid core growth} (\S~\ref{sec:solid_core_growth}) \\
    This component remains unchanged from KI22.
    \item \textbf{Envelope accumulation} (\S~\ref{sec:atmosphere}) \\
    The overall treatment is consistent with KI22, with minor updates to the equation of state and the modelling of planetary luminosity due to core cooling.
    \item \textbf{Post-disc thermal evolution and loss of envelopes}  (\S~\ref{sec:radius}) \\
    The envelope photoevaporation model has been substantially updated. We also revise the thermal evolution treatments for gas-rich and water-rich planets.
    \item \textbf{Orbital migration and resonance trapping} (\S~\ref{sec:orbital_evolution})\\
    Resonance trapping is now treated with an improved model, and tidal migration is newly incorporated.
    \item \textbf{Post-disc dynamical evolution} (\S~\ref{sec:dynamical_interact}) \\
    This component has been extensively revised by introducing a new semi-analytical model.
\end{itemize}

In addition, the initial conditions for planetary embryos have been updated from KI22, whereas the probability distribution of Monte Carlo variables, such as the initial disc gas mass and metallicity,
%those for the discs 
remain largely unchanged. Some parameter values, such as disc viscosity, have also been modified (\S~\ref{sec:setting}).

\subsection{Stellar Evolution} \label{sec:star}

The evolution of central stars influence the thermal evolution of surrounding protoplanetary disks.
We track the evolution of stellar radius ($R_*$), luminosity ($L_*$), and effective temperature ($T_*$) as functions of stellar mass ($M_*$) and age, using the tabulated data from \cite{Baraffe+2015}. This represents an update from KI22, which employed the dataset from \cite{Baraffe+1998}.

In addition to bolometric emission, high-energy radiation plays a crucial role in modelling the photoevaporation of disc gas and planetary atmospheres. As in KI22, we adopt the X-ray and EUV luminosities ($L_{\rm X}$ and $L_{\rm EUV}$) from the tables provided by \cite{Johnstone+2021}, which offer best-fit evolutionary tracks as functions of $M_*$ and the initial stellar rotation rate during the early phase ($\sim$150~Myr).
\rev{The model of \cite{Johnstone+2021} is based on the stellar evolution tracks of \cite{Spada+etal2013}, which were shown to be consistent with \cite{Baraffe+1998} (and thus with \cite{Baraffe+2015}). 
Therefore, we consider the bolometric and XUV luminosity evolution adopted here to be mutually consistent. 
}

\rev{
Note that the rotation model of \cite{Johnstone+2021} includes a disc-locking phase, during which the stellar rotation rate remains constant. The duration of this phase is not explicitly tied to the disc lifetime in our simulations; however, the resulting inconsistency has only a minor impact on the disc evolution.}

The initial stellar rotation rate, $\Omega_{*0}$, is treated as a Monte Carlo variable, and the corresponding evolutionary tracks of $L_{\rm X}$ and $L_{\rm EUV}$ are selected accordingly. To account for the observed scatter among stars, we introduce a deviation factor, following the approach adopted in KI22.

\subsection{Protoplanetary Disc} \label{sec:disc}
\subsubsection{Gas disc profile and evolution}

We model the evolution of the gas disc through viscous diffusion and photoevaporation, following the general framework of KI22. The updates introduced here are (i) modified initial profiles and (ii) a new prescription for photoevaporation.

\begin{center}
\textit{(i) Initial profiles}
\end{center}

The initial gas surface density is assumed to follow \citep{Veras+Armitage2004,Andrews+2010}:
\begin{align}
    \Sigma_{\rm g}^{(t=0)} &= \Sigma_{\rm g0}\qty(\frac{r}{r_0})^{-q_{\rm g}}
    \exp[-\qty(\frac{r}{r_{\rm disc}})^{2-q_{\rm g}}],
    \label{eq:Sigmag0}
\end{align}
where $\Sigma_{\rm g0}$ is the surface density at $r=r_0=1$~au, and $r_{\rm disc}$ is the characteristic disc radius. 
In KI22, we introduced an additional tapering factor to make $\Sigma_{\rm g}\rightarrow 0$ near the disc inner edge, following recent planet population synthesis studies~\citep[e.g.,][]{Brugger+2020,Emsenhuber+2021a}.
We found, however, that the formation of close-in planets is sensitive to the assumed tapering shape, which remains unconstrained due to the uncertain structure of the innermost disc region.  
To avoid introducing additional, model-dependent uncertainty, this factor is omitted in the present work.
A detailed investigation of how the treatment of the disc inner edge affects planet formation is an important topic for future work, but lies beyond the scope of this study.
%The difference from KI22 is that we no longer impose $\Sigma_{\rm g}\to 0$ near the disc inner edge.

As in KI22, we adopt $q_{\rm g}=0.9$, consistent with disc observations~\citep{Andrews+2010}, and fix $r_0=1$~au. 
The normalisation $\Sigma_{\rm g0}$ is set so that the integrated surface density between the inner edge $r_{\rm in}$ and $r_{\rm max}=1000$~au yields the total disc gas mass $M_{\rm disc}$. 
The inner edge $r_{\rm in}$ is treated as a Monte Carlo variable, defined as the orbital radius where the Keplerian period equals the stellar rotation period (see Section~\ref{sec:init_cond_disc}).

\begin{center}
\textit{(ii) Viscous diffusion}
\end{center}

The time evolution of the disc gas surface density $\Sigma_{\rm g}$ is described by the same diffusion equation as in KI22, including viscous diffusion, photoevaporation, and planetary accretion \citep[e.g.,][]{Lynden-Bell+Pringle1974}:
\begin{equation}
    \pdv{\Sigma_{\rm g}}{t} - \frac{3}{r}
    \pdv{r}\left( r^{1/2} \pdv{r}(\nu_{\rm acc} \Sigma_{\rm g} r^{1/2})\right)
    = -\dot{\Sigma}_{\rm pe} - \dot{\Sigma}_{\rm planet},
    \label{eq:dSigma_dt_basic}
\end{equation}
where $\dot{\Sigma}_{\rm pe}$ and $\dot{\Sigma}_{\rm planet}$ represent sink terms due to photoevaporation and planetary accretion, respectively. 
The treatment of $\dot{\Sigma}_{\rm planet}$ remains unchanged from KI22, whereas the photoevaporation model has been substantially revised (see below).

As in KI22, we use the $\alpha$-prescription for turbulent viscosity, 
$\nu_{\rm acc}=\alpha_{\rm acc} c_{\rm s}H_{\rm disc}$ \citep{Shakura+Sunyaev1973}, 
where $c_{\rm s}$ is the sound speed and $H_{\rm disc}$ the scale height. 
The viscosity parameter $\alpha_{\rm acc}$ is an input parameter.

Equation~\eqref{eq:dSigma_dt_basic} is solved on a logarithmic radial grid of $N_\mathrm{grid,disc}=500$ points between $r_{\rm in}$ and $r_{\rm max}$. 
We impose the standard zero-torque boundary condition at the inner edge, $\pdv*{(\nu_{\rm acc}\Sigma_{\rm g}r^{1/2})}{r}=0$ \citep{Lynden-Bell+Pringle1974}, and set $\Sigma_{\rm g}(r_{\rm max})=0$ at the outer boundary.
\rev{
For numerical stability, we impose a surface-density floor of $\Sigma_{\rm g}=10^{-10}~{\rm g\,cm^{-2}}$. 
The integration timestep is adaptively chosen such that the relative change in $\Sigma_{\rm g}$, the planetary semi-major axis, core mass, and envelope mass does not exceed 5\% per step.
}

\begin{center}
\textit{(iii) Photoevaporation}
\end{center}

In contrast to KI22, which considered EUV-driven internal and FUV-driven external photoevaporation, we now adopt an X-ray-driven model, because it is known to be more effective than EUV photoevaporation~\citep[e.g.,][]{Owen+2010,Picogna+etal2019,Ercolano+etal2021}. %. 
Specifically, we use the prescription of \cite{Owen+2012}, based on the radiation hydrodynamic simulations of photoevaporating discs done by \cite{Owen+2010,Owen+2011}. 
The mass-loss rate $\dot{M}_{\rm X}$ is given as a function of stellar X-ray luminosity $L_{\rm X}$ and stellar mass $M_{\rm X}$ by
\begin{equation}
    \dot{M}_{\rm X} = 6.25\times 10^{-9} f_{\rm PE}
    \qty(\frac{L_{\rm X}}{10^{30}~{\rm erg~s^{-1}}})^{1.14}
    \qty(\frac{M_*}{M_\odot})^{-0.068}
    ~M_\odot {\rm yr^{-1}},
    \label{eq:dM_PE_1}
\end{equation}
before inner hole opening, and
\begin{equation}
    \dot{M}_{\rm X} = 4.8\times 10^{-9}f_{\rm PE}
    \qty(\frac{L_{\rm X}}{10^{30}~{\rm erg~s^{-1}}})^{1.14}
    \qty(\frac{M_*}{M_\odot})^{-0.148}
    ~M_\odot{\rm yr^{-1}},
    \label{eq:dM_PE_2}
\end{equation}
afterwards. 
The radial profiles of $\dot{\Sigma}_{\rm pe}$ for both regimes are taken from \cite{Owen+2012}. 
\rev{
In this prescription, photoevaporation prior to inner-hole opening operates in regions satisfying
\begin{equation}
    x = 0.85 \left(\frac{r}{\rm au}\right)\left(\frac{M_*}{M_\odot}\right)^{-1} > 0.7.
\end{equation}
We therefore identify inner-hole opening when $\Sigma_{\rm g}$ reaches the numerical floor throughout the region $x<0.7$. 
Physically, the transition between Eqs.~\eqref{eq:dM_PE_1} and \eqref{eq:dM_PE_2} should occur once the radial X-ray optical depth in the inner disc falls below unity. However, because the remaining inner-disc gas is rapidly dispersed at this stage, the precise transition criterion has only a minor effect on the overall disc evolution.
}

We include a scaling factor $f_{\rm PE}$ to account for uncertainties in efficiency. 
Recent work incorporating detailed atomic and molecular cooling indicates that mass-loss rates for solar-type stars may be reduced by a factor of %$\sim$0.1–0.3
$\sim$3-10~\citep{Sellek+etal2024a}. 
Although the dependence on high-energy flux and stellar mass remains uncertain, we adopt $f_{\rm PE}=0.3$ uniformly for all stellar types.

\subsubsection{Disc temperature}\label{sec:disk_temp}

As in KI22, the midplane temperature $T_{\rm disc}$ is determined by the combined effects of viscous heating, indirect stellar irradiation, and direct stellar irradiation:
\begin{equation}
    T_{\rm disc}^4 = T_{\rm vis}^4 + T_{\rm irr}^4 + T_{\rm eq}^4 \exp(-\tau_{r}),
\end{equation}
where $T_{\rm vis}$ and $T_{\rm irr}$ denote the contributions from viscous heating and indirect irradiation, respectively, $T_{\rm eq}$ is the equilibrium temperature from direct irradiation, and $\tau_r$ is the radial \rev{IR} optical depth at the disc midplane\rev{, computed from the gas opacity of \cite{Freeman+2014} and the grain opacity of \cite{Semenov+2003}. 
Although this differs from the optical depth relevant for stellar visible photons, we adopt this approximation for simplicity, as the $\exp(-\tau_r)$ term becomes important only during the final dispersal stage of the disc.} 
In KI22, $T_{\rm vis}$ and $T_{\rm irr}$ were computed iteratively, since both depend on opacity and scale height, which in turn depend on $T_{\rm disc}$. 
Here, to reduce computational cost, we adopt simplified but physically motivated prescriptions that reproduce the main features of the temperature profile, as outlined below.

In an optically thick disc, the viscous heating contribution is given by \citep{Nakamoto+Nakagawa1994}:
\begin{equation}
    \sigma T_{\rm vis,NN}^4 = \frac{27}{64}\,\kappa\, \nu_{\rm mid}\, \Sigma_{\rm g}^2\,\Omega_{\rm K}^2,
    \label{eq:Tvis_irr}
\end{equation}
where $\nu_{\rm mid}=\alpha_{\rm mid} c_{\rm s}H_{\rm disc}$ is the midplane viscosity.  
As in KI22, we distinguish the midplane turbulence parameter $\alpha_{\rm mid}$ from the accretion parameter $\alpha_{\rm acc}$.

We adopt a constant dust opacity of $\kappa = 5~{\rm cm^2~g^{-1}}$, which is a typical value for silicate dust inside the water iceline in solar-abundance discs,
where viscous heating dominates~\citep{Semenov+2003}.  
At $T_{\rm disc}>T_{\rm evap}\simeq 1500$~K, dust sublimation causes a sharp opacity drop, making $T_{\rm vis}$ nearly independent of $r$ \citep[e.g.,][]{DAlessio+2001,Hueso+Guillot2005}.  
To capture this behaviour smoothly, we use the approximation
\begin{equation}
    T_{\rm vis}^{-n} = T_{\rm evap}^{-n} + T_{\rm vis,NN}^{-n},
\end{equation}
with $n=8$, which ensures a steep but continuous transition. 
The precise choice of $n$ has negligible influence on disc or planet evolution.

The contribution from stellar irradiation at the disc surface is \citep{Kusaka+1970,Adams+1988,Ruden+Pollack1991}:
\begin{align}
    T_{\rm irr,org}^4 &= T_*^4 
    \left[ \frac{2}{3\pi}\qty(\frac{R_*}{r})^3 
    + \frac{1}{2}\qty(\frac{R_*}{r})^2
    \frac{H_{\rm disc}}{r}\left(\dv{\ln H_{\rm disc}}{\ln r}-1\right)\right].
    \label{eq:Tirr_org}
\end{align}
The first term dominates in the inner disc region (where $H_{\rm disc}<R_*$), and the second in the outer region ($H_{\rm disc}>R_*$).  
We approximate these two regimes by
\begin{align}
    T_{\rm irr,in} &= T_* \qty(\tfrac{2}{3\pi})^{1/4} \qty(\tfrac{R_*}{r})^{3/4}, \\
    T_{\rm irr,out} &= \qty[
    \tfrac{9}{14}\sqrt{\tfrac{k_{\rm B}}{\mu_{\rm disc}m_{\rm H}}}
    \frac{T_*^4R_*^2}{r^3 \Omega_{\rm K}}
    ]^{2/7},
\end{align}
where $T_{\rm irr,out}$ is derived from the second term of Eq.~\eqref{eq:Tirr_org}, assuming $\dv*{\ln H_{\rm disc}}{\ln r}=9/7$ \citep{Chiang+Goldreich1997}.  
The combined irradiation temperature is then
\begin{equation}
    T_{\rm irr}^4 = T_{\rm irr,in}^4 + T_{\rm irr,out}^4.
\end{equation}

Finally, the equilibrium temperature due to direct stellar irradiation is
\begin{equation}
    T_{\rm eq}^4 = \frac{L_*}{16\pi \sigma r^2}.
    \label{eq:Teq}
\end{equation}

\subsubsection{Distribution and evolution of planetesimals}

The initial planetesimal surface density is prescribed as
\begin{equation}
    \Sigma_{\rm s}^{(t=0)} = \eta_{\rm ice}\,\Sigma_{\rm s0}
    \qty(\frac{r}{r_0})^{-q_{\rm s}}
     \exp\!\left[-\qty(\frac{r}{r_{\rm solid}})^2\right],
    \label{eq:Sigma_s}
\end{equation}
where $\Sigma_{\rm s0}$ is the surface density at $r=r_0$, $r_{\rm solid}$ is the characteristic radius introduced to take into account the relatively sharp outer edges of the observed dust discs~\citep{Birnstiel+Andrews2014}, and $\eta_{\rm ice}$ accounts for H$_2$O condensation and sublimation.  
This profile and its parameter values are the same as in KI22, except that, as for $\Sigma_{\rm g}^{(t=0)}$, we no longer apply a smooth truncation near the disc inner edge.  
Instead, to prevent planetesimals from existing in extremely hot regions, we impose $\Sigma_{\rm s}=0$ whenever $T_{\rm disc} \ge 0.99T_{\rm evap}$.  
The factor 0.99 is introduced because, under our temperature prescription, $T_{\rm disc}$ does not strictly reach $T_{\rm evap}$.

Each planetesimal is assigned a bulk density $\rho_{\rm plt}$ of $3.2~{\rm g~cm^{-3}}$ inside the iceline and $1~{\rm g~cm^{-3}}$ outside.  
Unlike KI22, where all planetesimals had the same mass, here we assume a common radius $R_{\rm plt}$ for all planetesimals, specified as an input parameter, and compute their masses $M_{\rm plt}$ from $\rho_{\rm plt}$ and $R_{\rm plt}$.  

The planetesimal surface density evolves through accretion and scattering by protoplanets, calculated in the same manner as in KI22.

\subsection{Solid Core Growth} \label{sec:solid_core_growth}

The growth of planetary cores by planetesimal accretion is modelled following KI22, which adopted the prescriptions of \citet{Inaba+2001} with the capture radius $R_{\rm cap}$ defined by \citet{Inaba+Ikoma2003}.  
The core mass growth rate is
\begin{equation}
    \dv{M_{\rm core}}{t} = \Omega_{\rm K}\,\bar{\Sigma}_{\rm s}\,R_{\rm H}^2\,p_{\rm col},
    \label{eq:dMc_dt}
\end{equation}
where $\bar{\Sigma}_{\rm s}$ is the mean planetesimal surface density in the feeding zone and $p_{\rm col}$ is the collision probability, dependent on planetesimal eccentricity $e_{\rm plt}$ and inclination $i_{\rm plt}$ (see \citealt{Inaba+2001} for the explicit form).  
We adopt $i_{\rm plt}=e_{\rm plt}/2$.  

\rev{
The planetesimal capture radius $R_{\rm cap}$ is calculated by
\begin{equation}
    R_{\rm plt} = \frac{3}{2}
    \frac{v_{\rm ran}^2+2GM_{\rm p}/R_{\rm cap}}{v_{\rm ran}^2+2GM_{\rm p}/R_{\rm H}}
    \frac{\rho (R_{\rm cap})}{\rho_{\rm plt}}R_{\rm cap},
\end{equation}
where $v_{\rm ran} = e_{\rm plt}a_{\rm p}\Omega_{\rm K}$ is the planetesimal random velocity and $\rho(R_{\rm cap})$ is the envelope gas density at radius $R_{\rm cap}$.
}

The eccentricity $e_{\rm plt}$ in the feeding zone is set by the balance between viscous stirring by the protoplanet and damping by gas drag \citep{Kokubo+Ida2002}.  
As in KI22, it is expressed as
\begin{equation}
    e_{\rm plt} =
    \qty(\frac{12\ln \Lambda}{C_{\rm D}\,2^{1/3}\,b_{\rm FZ}}
    \frac{\rho_{\rm plt}R_{\rm plt}}{\rho_{\rm disc}a_{\rm p}})^{1/5}
    \frac{R_{\rm H}}{a_{\rm p}},
    \label{eq:e_plt}
\end{equation}
where $C_{\rm D}$ is the gas drag coefficient, $b_{\rm FZ}$ is the feeding zone width in mutual Hill radii, and $\ln \Lambda$ represents the effect of repeated distant encounters.  
We adopt $C_{\rm D}=1$, $b_{\rm FZ}=10$, and $\ln \Lambda=3$ \citep{Kokubo+Ida2002}.  

After disc gas dispersal, $e_{\rm plt}$ is assumed to be limited by the escape eccentricity,
\begin{equation}
    e_{\rm esc} = \frac{v_{\rm esc}}{v_{\rm K}} = 
    \sqrt{\frac{2GM_{\rm p}/R_{\rm cap}}{GM_*/a_{\rm p}}},
\end{equation}
where $v_{\rm esc}$ and $v_{\rm K}$ are the planetary escape and Keplerian velocities, respectively \citep{Safronov1969,Kokubo+Ida2002}.  

Once the gas is dissipated, planetesimals encountering a planet may be ejected rather than accreted.  
Following \citet{Ida+Lin2004} and \citet{Emsenhuber+2021a}, the ejection rate is estimated by comparing collision and scattering cross-sections.  
Assuming that planetesimals are ejected when their relative velocity exceeds $v_{\rm K}$, the ejection rate is
\begin{equation}
    \dot{M}_{\rm scat} = e_{\rm esc}^4\,\dot{M}_{\rm core}.
\end{equation}
This prescription is used in calculating the planetesimal surface density evolution.

\subsection{Envelope Accumulation} \label{sec:atmosphere}
\subsubsection{Overview of Envelope Model}

The formation and evolution of the primordial atmosphere (hereafter the envelope) are modelled following KI22, but with several updates described below. We divide the process into four phases (I–IV):

\begin{itemize}
    \item \textbf{Phase I: Hydrostatic and Thermally Steady Envelope} \\
    During planetesimal accretion, the envelope remains hydrostatic and in thermal equilibrium. The temperature at the bottom of the envelope is high enough to maintain a global magma ocean~\citep[][]{Ikoma+Genda2006,Kimura+Ikoma2020}. Interaction between the envelope and the magma ocean produces volatiles, significantly altering the envelope composition. 
    \rev{Since the accretion luminosity is high and the envelope is largely convective at this stage,}
    we assume that water-producing reactions are efficient and that the resulting vapour is uniformly mixed. We refer to this state as the \textit{vapour-mixed envelope}, with mass $M_{\rm mix}$ and water mass fraction $X_{\rm H_2O,mix}$, the latter treated as an input parameter.

    \item \textbf{Phase II: Envelope Contraction and Disc Gas Accumulation} \\
    Once planetesimal accretion declines, the vapour-mixed envelope contracts gravitationally and accretes additional disc gas, primarily H and He. We assume that this newly accreted gas does not mix with the vapour-mixed layer, so $M_{\rm mix}$ remains fixed. The accumulated disc gas mass is denoted $M_{\rm HHe}$. We compute the quasi-static contraction of this two-layer envelope.
    \rev{
    This represents a limiting case in which a strong compositional gradient between the vapour-mixed and H–He layers suppresses convective mixing of these layers. The implications of this assumption are discussed in \S~\ref{sec:caveats}.
    }
    \item \textbf{Phase III: Runaway Gas Accretion} \\
    The planet enters runaway gas accretion. As in Phase II, the vapour-mixed envelope does not mix with the incoming H–He gas. We therefore fix $M_{\rm mix}$ and compute the evolution of $M_{\rm HHe}$ using empirical formulae without solving the full envelope structure.

    \item \textbf{Phase IV: Long-Term Thermal Evolution and Escape} \\
    After disc dispersal, the envelope undergoes long-term cooling and atmospheric escape. Details of this phase are given in \S~\ref{sec:radius}.
\end{itemize}

Below, we describe the models for Phases I--III, highlighting differences from KI22.

\subsubsection{Phase I: Purely Hydrostatic Equilibrium}
\label{sec:hydrostatic_atmosphere}

We use the 1D internal structure model of \citet{Kimura+Ikoma2020}, which solves the standard stellar structure equations. The overall treatment is the same as in KI22, but with updated equations of state (EOS).

The envelope consists of H, He, and O. We adopt the H–He EOS of \citet{Chabrier+2021} and the H$_2$O EOS of \citet{Haldemann+2020}. The latent heat released by H$_2$O condensation is included in the adiabatic temperature gradient. 
\rev{
We assume that condensed H$_2$O rains out and does not contribute to the local density or heat capacity. The temperature gradient in the condensation region therefore follows a pseudo-moist adiabat~\citep{Kasting1988}.
For simplicity, the condensate is not added to deeper layers or to the core. The resulting mass inconsistency is negligible, as the condensation region contains only a small fraction of the total envelope mass. 
Neglecting precipitation generally yields larger envelope masses~\citep{Venturini+2015,Kimura+Ikoma2020}; our treatment therefore likely provides a conservative estimate of the impact of water enrichment in cold environments. See \citet{Kimura+Ikoma2020} for further discussion.
}
For opacities, we use the gas opacity table of \citet{Kimura+Ikoma2020} and the dust opacity of \citet{Semenov+2003}, with a dust depletion factor from \citet{Ormel2014}. See \citet{Kimura+Ikoma2020} for details.

The envelope mass $M_\mathrm{mix}$ is computed with $X_{\rm H_2O,mix}$ as an input parameter. The outer boundary is set to the minimum of the Bondi and Hill radii, with pressure and temperature taken from the disc values there. The inner boundary is the solid core radius $R_{\rm core}$, calculated from the fitting formulae of \citet{Fortney+2007} and \citet{Zeng+2019}.

The luminosity $L$ is assumed constant throughout the envelope and equal to the core luminosity $L_{\rm core}$, which is the sum of contributions from planetesimal accretion ($L_{\rm acc}$) and radioactive decay ($L_{\rm radio}$), as in KI22.

\subsubsection{Phase II: Quasi-Static Contraction}
\label{sec:quasi_static_ppsm}

As planetesimal accretion rate declines, the envelope contracts quasi-statically and accretes disc gas on top of the vapour-mixed layer. This phase follows KI22, based on the energy conservation approximation~\citep{Papaloizou+Nelson2005,Mordasini+2012b,Fortier+2013,Piso+Youdin2014,Venturini+2016}, but with two updates: (i) the treatment of core cooling luminosity, and (ii) the criterion for transition from Phase I.

Energy conservation between $t-\Delta t$ and $t$ gives
\begin{align}
    \frac{E_{\rm env}(t)-E_{\rm env}(t-\Delta t)}{\Delta t} = L_{\rm core}
    + e_{\rm gas} \frac{M_{\rm env}(t) - M_{\rm env}(t-\Delta t)}{\Delta t} - L,
    \label{eq:energy_conserv}
\end{align}
where $e_{\rm gas}$ is the total specific energy (internal + gravitational) of the disc gas at the outer boundary, $M_{\rm env}=M_{\rm mix}+M_{\rm HHe}$ is the total envelope mass, and $E_{\rm env}$ is the total envelope energy defined by
\begin{equation}
    E_{\rm env} = \int_{M_{\rm core}}^{M_{\rm p}} \qty( u - \frac{GM_R}{R}) \dd{M_R}.
\end{equation}
Here $u$ is the specific internal energy and $M_R$ is the total mass enclosed within a sphere of radius $R$.

The core luminosity is
\begin{equation}
    L_{\rm core}=L_{\rm acc} + L_{\rm cool} + L_{\rm radio},
    \label{eq:Lcore_II}
\end{equation}
with $L_{\rm cool}$ from core cooling. Unlike KI22, which used an analytical approximation, we compute $L_{\rm cool}$ explicitly as
\begin{equation}
    L_{\rm cool} = -M_{\rm core} C_{\rm rock}\dv{T_{\rm surf}}{t}
    = -M_{\rm core} C_{\rm rock}\frac{T_{\rm surf}(t) - T_{\rm surf}(t-\Delta t)}{\Delta t},
    \label{eq:Lcool}
\end{equation}
where $C_{\rm rock}=1.2\times 10^7~{\rm erg~g^{-1}K^{-1}}$ is the rock specific heat, and $T_{\rm surf}$ the core surface temperature.

Substituting into Eq.~\eqref{eq:energy_conserv} gives
\begin{equation}
    \Delta t = 
        \frac{\Delta E}
        {L_\mathrm{acc} + L_{\rm radio} - L},
    \label{eq:energy_conserv2}
\end{equation}
with
\begin{align}
    \Delta E &= E_{\rm env}(t)-E_{\rm env}(t-\Delta t) \notag \\
    &\quad + M_{\rm core} C_{\rm rock}\qty[ T_{\rm surf}(t) - T_{\rm surf}(t-\Delta t)] \notag \\
    &\quad -e_{\rm gas} \left[M_{\rm env}(t) - M_{\rm env}(t-\Delta t)\right].
\end{align}
We first assume $L$, integrate the structure equations to obtain \rev{the temporal values of $E_{\rm env}(t)$ and $M_{\rm env}(t)$}, then compute $\Delta t$ using Eq.~\eqref{eq:energy_conserv2}. This procedure is iterated until $\Delta t$ matches the system timestep (see KI22).
\rev{
Here the H$_2$O fraction is set to zero for $M_r > M_{\rm core} + M_{\rm mix}$ and to $X_{\rm H_2O,mix}$ otherwise.
}

\rev{
The transition between Phase~I and Phase~II is determined by whether the envelope remains in thermal equilibrium or undergoes contraction.
Phase~I corresponds to a purely hydrostatic and thermally steady state, in which the envelope luminosity equals the core luminosity,
\begin{equation}
L = L_{\rm acc}(t) + L_{\rm radio}(t).
\end{equation}
In contrast, Phase~II begins when the envelope luminosity exceeds the core luminosity and gravitational contraction contributes to the energy budget.
}

\rev{
Numerically, at each timestep we evaluate the cooling timescale required for the envelope luminosity to decline from its previous value $L(t-\Delta t)$ to the current core luminosity $L_{\rm acc}(t)+L_{\rm radio}(t)$, using Eq.~\eqref{eq:energy_conserv2}.
If this cooling timescale is shorter than the system timestep $\Delta t_{\rm sys}$, the envelope has already relaxed to thermal equilibrium and remains in Phase~I. 
If it is longer than $\Delta t_{\rm sys}$, the envelope has not yet equilibrated and continues contraction, defining Phase~II.
In practice, the transition typically occurs shortly after the solid accretion rate declines and $L_{\rm acc}$ decreases from its peak value. 
If solid accretion resumes, for example due to inward migration into a swarm of planetesimals, the core luminosity can increase again and the planet can revert to Phase~I.
Note that the transition to Phase II also happens when the hydrostatic solution is not found in Phase I (i.e., the critical core mass for the given $X_{\rm H_2O,mix}$ is reached).
}

Finally, once the critical luminosity, below which no hydrostatic solution exists~\citep[][]{Ikoma+2000,Hubickyj+2005,Ikoma+25}, is reached, the planet undergoes runaway gas accretion (Phase III).

\subsubsection{Phase III: Runaway Gas Accretion Phase}
\label{sec:runaway_gas_acc}

The model for Phase III is identical to that used in KI22. In this phase, we no longer solve the envelope structure explicitly. Instead, the envelope mass $M_{\rm HHe}$ is increased according to the runaway gas accretion rate:
\begin{equation}
    \dv{M_{\rm HHe}}{t} = \min\qty(\dot{M}_{\rm KH}, \dot{M}_{\rm gap}, \dot{M}_{\rm disc}),
    \label{eq:gas_acc_rate}
\end{equation}
where $\dot{M}_{\rm KH}$ is the rate governed by Kelvin–Helmholtz contraction, evaluated using the timescale prescription from \cite{Ikoma+2000}.
The second term, $\dot{M}_{\rm gap}$, represents the gas supply rate through the disc gap formed by disc–planet interaction, and is calculated following the prescription of \cite{Tanigawa+Ikoma2007}, with the gas surface density at the gap bottom given by \cite{Kanagawa+2015}.
\rev{Note that the mid-plane viscosity $\alpha_{\rm mid}$ is used to calculate the gap depth.}
Finally, $\dot{M}_{\rm disc}$ is the global disc accretion rate, determined from the disc gas evolution model.
For the explicit forms of these accretion rates, see KI22.

\subsection{Envelope Thermal Evolution and Loss (Phase IV)}
\label{sec:radius}

After disc dispersal, the planetary radius and envelope mass evolve through photoevaporation and thermal cooling (Phase IV). 
Both the photoevaporation and thermal evolution models are updated relative to KI22.

We assume that the envelope detaches from the disc gas once either of the following conditions is satisfied: (i) $\rho_{\rm disc} < 10^{-18}~{\rm g/cm^3}$, or (ii) $\tau_{r} < 1$. 
The first condition approximately corresponds to the case where the mean free path of the disc gas becomes comparable to an Earth-sized planetary radius.
The second condition represents the stage at which stellar radiation can directly reach the planet, initiating envelope photoevaporation.  
Although the optical depth for stellar XUV is generally larger than $\tau_r$ (which is the IR optical thickness), the difference is unimportant because the disc disperses rapidly once it becomes optically thin ($\tau_r < 1$).

As in KI22, the treatment of thermal evolution depends on whether the planet has undergone runaway gas accretion. 
In particular, the thermal model for planets with thick envelopes (such gas giants) has been significantly revised.  
Additionally, once the entire envelope has escaped, we compute the thickness of any remaining H$_2$O layer by solving its internal structure, in order to capture the radii of close-in planets with hot steam envelopes~\citep{Burn+etal2024}.

\subsubsection{Thermal Evolution of Thin Envelopes}
\label{sec:evolution_thin_envelope}

For planets that have not undergone runaway gas accretion, we compute the thermal evolution of the envelopes using the same method as in \S~\ref{sec:quasi_static_ppsm}, deriving the planetary radius $R_{\rm p}$ (defined at the 10~mbar level). 
This treatment remains unchanged from KI22.

For numerical convenience, we separate the upper atmosphere from the deeper envelope and calculate the radiative–convective structure of the atmosphere to provide the outer boundary conditions for the envelope. See KI22 for details of the procedure.

To reduce computational cost, we use a precomputed table that relates
\begin{equation}
    (X_{\rm H_2O}, T_{\rm eq}, M_{\rm core}, M_{\rm mix}, M_{\rm HHe}, L) \to (R_{\rm p}, T_{\rm surf}, E_{\rm env}),
\end{equation}
and then evaluate Eq.~\eqref{eq:energy_conserv2} by interpolation.
\rev{
See Appendix~\ref{sec:grid_thin} for the details of the table structure and comparison with direct calculation.
}

\subsubsection{Thermal Evolution of Thick Envelopes}
\label{sec:evolution_thick_envelope}

If a planet has experienced runaway gas accretion, it acquires a thick H–He envelope of nearly nebular composition.  
Whereas KI22 used the empirical radius formula of \citet{Lopez+Fortney2014}, here we directly simulate the thermal evolution.

The evolution of such planets (including gas giants) cannot be described by Eq.~\eqref{eq:energy_conserv2}, as the assumption of constant luminosity within the envelope no longer holds. 
Instead, we assume the envelope is fully convective, so the energy conservation equation becomes
\begin{equation}
    L - L_{\rm core} = -\int_{M_{\rm core}}^{M_{\rm p}} T(t) \dv{S}{t} \dd{M_{R}}
     = -\dv{S_{\rm env}}{t} \int_{M_{\rm core}}^{M_{\rm p}} T(t) \dd{M_{R}},
     \label{eq:int_dSdm_constS}
\end{equation}
where $S_{\rm env}$ is the envelope specific entropy, assumed uniform. % \ikoma{and $M_R$ is the total mass enclosed within a sphere of planet-centric radius $R$}.

Substituting $L_{\rm core} = L_{\rm cool} + L_{\rm radio}$ (with $L_{\rm cool}$ from Eq.~\eqref{eq:Lcool}), the timestep is obtained from Eq.~\eqref{eq:int_dSdm_constS} as~\citep{Kurosaki+Ikoma2017}
\begin{equation}
    \Delta t = \frac{\Delta E}{L(t)+L(t-\Delta t) -2L_{\rm radio}},
\end{equation}
with
\begin{align}
    \Delta E &= -2C_{\rm rock}M_{\rm core}[T_{\rm surf}(t)-T_{\rm surf}(t-\Delta t)] \notag \\
    &\quad - [S_{\rm env}(t)-S_{\rm env}(t-\Delta t)][\Theta(t)+\Theta (t-\Delta t)],
\end{align}
and
\begin{equation}
    \Theta (t) = \int_{M_{\rm core}}^{M_{\rm p}} T(t) \dd{M_{R}}.
\end{equation}

The outer boundary conditions are determined as in \S~\ref{sec:evolution_thin_envelope}.
However, since the envelope is assumed to be fully adiabatic, the atmosphere–envelope boundary must lie below the tropopause, with the atmospheric mass and thickness negligible compared to those of the envelope. 
We thus set $P_{\rm out}=1$~kbar, which satisfies these conditions for most close-in planets.

To further reduce computational cost, we performed grid simulations of thermal evolution from $t=10^6$ to $10^{10}$~yr with sufficiently high initial entropy, and constructed a table relating
\begin{equation}
    (T_{\rm eq}, M_{\rm core}, M_{\rm env}, t) \to (R_{\rm p}, T_{\rm surf}).
\end{equation}
Since we do not solve the detailed structure during the runaway accretion phase itself, the post-disc initial thermal state is not explicitly computed. Instead, we interpolate the above table at the system age $t$ to obtain $R_{\rm p}$. 
Because long-term cooling is largely insensitive to the initial entropy~\citep[e.g.][]{Marley+etal2007}, this simplification has negligible impact on the final planetary radius.
\rev{
More details on the table are described in Appendix~\ref{sec:grid_thick}.
}

\subsubsection{Thermal Evolution of the Water Layer}
\label{sec:evolution_water_layer}
Planets that have accreted large amounts of icy planetesimals have a thick H$_2$O layer above their rocky cores. 
In particular, close-in planets with high water fractions can retain a steam-rich atmosphere after the primordial envelope has been lost, which substantially inflates their radii~\citep{Burn+etal2024}.

To capture this effect, we newly implement the thermal evolution of the water layer in the absence of an envelope. 
The method follows that described in \S~\ref{sec:evolution_thick_envelope}, and we construct a grid mapping
\begin{equation}
    (T_{\rm eq}, M_{\rm rock}, M_{\rm ice}, t) \to (R_{\rm p}, T_{\rm surf}),
\end{equation}
where $M_{\rm rock}$ and $M_{\rm ice}$ denote the masses of rock and H$_2$O ice in the core, respectively. 
Once the planet has completely lost its envelope, we determine $R_{\rm p}$ by interpolating this table.
\rev{
Details of the grid construction are provided in Appendix~\ref{sec:grid_water}. 
Because the water layer is treated as fully condensed (following \cite{Zeng+2019}) while the envelope is present, a small radius discontinuity arises at the moment of complete envelope loss. The water-layer thickness during the envelope phase is typically underestimated by $\sim$10--20\%. 
However, since the H–He envelope dominates the planetary radius in this regime, the impact on the total radius and subsequent thermal evolution is minor. 
The discrepancy becomes relevant only when the envelope mass is quite small and comparable to the water-layer correction, although this phase is short-lived because such thin envelopes are rapidly removed.
}

\subsubsection{Atmospheric Photoevaporation}

After disc dispersal, envelope escape is driven by stellar XUV irradiation.  
While KI22 applied the same mass-loss prescription regardless of envelope composition, here we adopt different formulae for H–He and vapour-mixed envelopes.

For H–He envelopes, we use the fitting formula of \citet{Kubyshkina+2018b,Kubyshkina+2018a}, which predicts higher escape rates than the energy-limited approximation when the escape parameter is small ($\lesssim 10$):
\begin{equation}
    \dot{M}_{\rm esc,HHe} = e^\beta 
    \qty(\frac{F_{\rm XUV}}{{\rm erg~cm^{-2}~s^{-1}}})^{\alpha_1}
    \qty(\frac{a_{\rm p}}{{\rm au}})^{\alpha_2}
    \qty(\frac{R_{\rm p}}{R_\oplus})^{\alpha_3}
    \Lambda^k,
    \label{eq:escape_rate_kuby}
\end{equation}
where $F_{\rm XUV}$ is the incident XUV flux, $\Lambda$ is the escape parameter for atomic hydrogen, and $\beta$, $\alpha_1$, $\alpha_2$, $\alpha_3$, and $k$ are fitting coefficients depending on $\Lambda$ (see \citealt{Kubyshkina+2018a}).

In contrast, the hydrodynamic escape of water-rich envelopes is less efficient, owing to strong radiative cooling by H$_2$O and related chemical products~\citep{Yoshida+2022}.  
We therefore adopt the fitting formula of \citet{Yoshida+Gaidos2025} for vapour-mixed envelopes (and water layers):
\begin{equation}
    \dot{M}_{\rm esc,mix} = \dot{M}_{\rm ref} 
    \qty(\frac{F_{\rm EUV}}{10^3F_{\rm EUV}^\oplus})^a
    \qty(\frac{g}{4~{\rm m/s^2}})^{-3/2}
    \qty(\frac{M_{\rm p}}{5M_\oplus})^{1/2}.
\end{equation}
Here $F_{\rm EUV}$ is the stellar EUV flux, and $\dot{M}_{\rm ref}$ and $a$ depend on the H$_2$O/H$_2$ number ratio $r_{\rm H_2O}$ and $F_{\rm EUV}$ (see \citealt{Yoshida+Gaidos2025}).

We set $r_{\rm H_2O}=0.1$ as the maximum value, since the fitting formula is not calibrated beyond this value.  
At larger ratios, the upper atmosphere is expected to become optically thick in the infrared, saturating radiative cooling~\citep{Yoshida+Gaidos2025}.

It should be noted that we neglect fractionation between H$_2$ and H$_2$O molecules and assume that the escaping vapour-mixed envelope retains its bulk composition $X_{\rm H_2O,mix}$.  
This approximation is likely valid for close-in planets under strong irradiation, but for more temperate planets with Earth-like insolation, fractionation may prolong atmospheric survival~\citep{Yoshida+Gaidos2025}. 
This issue will be addressed in future work.

\subsection{Orbital Evolution}
\label{sec:orbital_evolution}

Planetary orbits evolve through disc-driven migration (type I or type II; \S~\ref{sec:migration}), resonance trapping of migrating pairs (\S~\ref{sec:resonance}), and subsequent resonant-chain dynamics near the disc inner edge (\S~\ref{sec:leakage}). After disc dispersal, we also account for orbital decay by star–planet tides (\S~\ref{sec:tide}).

\subsubsection{Type I and Type II Migration}
\label{sec:migration}

Orbital migration is treated in the same way as in KI22, by combining the type I migration prescriptions of \citet{Jimenez+Masset2017} and \citet{Masset2017} with the type II migration model of \citet{Kanagawa+2018}. The migration rate is given by
\begin{equation}
    \dv{a_{\rm p}}{t} = \frac{2\Gamma}{M_{\rm p} a_{\rm p} \Omega_{\rm K}},
    \label{eq:da_dt}
\end{equation}
where $\Gamma$ is the total torque acting on the planet, including the Lindblad, corotation, and thermal components, as well as the effects of gap opening. \rev{Note that the mid-plane viscosity $\alpha_{\rm mid}$ is used to calculate these torques and gap depth.}
See KI22 for details.

\subsubsection{Resonance Trapping}
\label{sec:resonance}

When planets undergo convergent migration, they are captured into mean-motion resonances~\citep{Murray+Dermott1999}, if the following conditions are fulfilled.  
Unlike KI22, here we newly implement resonance trapping by adopting the analytical criteria of \citet{Lin+etal2025}, which specify three conditions for capture:
\begin{enumerate}
    \item[(a)] The relative migration timescale must be long enough that the eccentricity excited by resonance remains below the maximum libration amplitude.
    \item[(b)] Gas damping of eccentricity must be weak enough to avoid significant changes in the semi-major axis that could lead to escape from resonance.
    \item[(c)] The resonance must be dynamically stable over long timescales.
\end{enumerate}

Here we consider only first-order resonances. The numerical procedure is:
\begin{enumerate}
    \item Identify a convergently migrating pair $(k, l)$, with $a_k < a_l$, and check whether they cross a $j:j-1$ resonance.  
    \item Compute the relative migration timescale
    \begin{equation}
        \frac{1}{\tau_{{\rm mig},kl}} = \frac{1}{\tau_{{\rm mig},l}} - \frac{1}{\tau_{{\rm mig},k}},
    \end{equation}
    with $\tau_{{\rm mig},i} = a_i/|\dot{a}_i|$.  
    The eccentricity damping timescale is given by \citep{Tanaka+Ward2004}
    \begin{equation}
        \tau_{\rm damp} = %\frac{1}{0.78}
        1.3\qty(\frac{M_{\rm p}}{M_*})^{-1}
        \qty(\frac{\Sigma_{\rm g}a_{\rm p}^2}{M_*})^{-1}
        \qty(\frac{H_{\rm disc}}{a_{\rm p}})^4 \Omega_{\rm K}^{-1},
        \label{eq:tau_dampTW}
    \end{equation}
    and we set $\tau_{{\rm damp},kl} = \min(\tau_{{\rm damp},k}, \tau_{{\rm damp},l})$.
    \item Evaluate the trapping criteria (a)--(c). See \cite{Lin+etal2025} for the explicit equations.  These equations use the disturbing function coefficients ($f_{\rm d,i}$ and $f_{\rm d,o}$ in \cite{Lin+etal2025}), which depend on Laplace coefficients~\citep{Murray+Dermott1999}.  
    For computational efficiency we approximate these as~\citep[e.g.][]{Goldberg+2022}:
    \begin{align}
        f_{\rm d,i} &= 0.4 - 0.8j, &
        f_{\rm d,o} &= 0.1 + 0.8j.
    \end{align}
    \item If all the conditions are satisfied, the planets are captured into $j:j-1$ resonance. Otherwise, convergent migration continues. If their separation shrinks below $2\sqrt{3}$ mutual Hill radii, they are assumed to collide and merge.
\end{enumerate}

Once trapped, the pair migrates while maintaining a fixed period ratio. The migration rate of such resonant pairs is calculated as in KI22, by redistributing their individual disc torques so as to preserve their semi-major axis ratio.  

\subsubsection{Leakage of Planets into the Disc Cavity}
\label{sec:leakage}

Migration usually halts at the disc inner edge. Subsequent planets migrating inward are then captured into resonances, forming a resonance chain. If the innermost planet has non-zero eccentricity such that part of its orbit extends into the gas disc, and its orbital velocity at aphelion is lower than that of the surrounding gas, it experiences a positive torque, which is called ``edge torque''~\citep{Ogihara+2010}.
This torque can counterbalance the negative torque on the planets in the resonance chain, thereby stalling their migration at the disc edge.

As in KI22, we model this by comparing the positive edge torque on the innermost planet with the total Type~I/II torque, $\Gamma$ (\S~\ref{sec:migration}), acting on the planets in the chain.
The eccentricity $e_1$ of the innermost planet, which is needed to evaluate the edge torque, is set to the equilibrium value from the balance between resonant excitation and gas damping \citep{Lin+etal2025}.  

If the edge torque is insufficient, the innermost planet is pushed into the cavity. We then assume that the semi-major axis ratio between this planet and its nearest neighbour remains constant.
%until orbital crossing or a giant impact occurs.

\subsubsection{Tidal Migration}
\label{sec:tide}

We newly include inward orbital decay due to star–planet tidal interactions.  
Following \citet{Emsenhuber+2021a}, we adopt a simplified prescription~\citep[e.g.][]{Jackson+etal2008,Ferraz-Mello+2008}:
\begin{equation}
    \dv{a_{\rm p}}{t} = -\frac{9}{2}\sqrt{\frac{G}{M_*}}
    \frac{R_*^5 M_{\rm p}}{Q_*}a_{\rm p}^{-11/2},
\end{equation}
where $Q_* = 10^6$ is the stellar tidal dissipation parameter. We account only for tides raised on the central star, neglecting those due to planetary deformation and eccentricity evolution.

This simplified treatment primarily serves to capture the engulfment of ultra-short-period planets ($\lesssim 1$~day) by their host star.
We should note that the expression above, together with the assumption of a constant $Q_*$, assumes that the planet orbits well inside the stellar corotation radius, where the orbital frequency exceeds the stellar spin frequency and the stellar rotation rate has only a minor effect on $Q_*$.
Indeed, since $\dv*{a_{\rm p}}{t}$ depends quite steeply on $a_{\rm p}$, the tidal effect becomes significant only for planets inside the disc inner edge, which approximately coincides with the corotation radius in the planet formation stage.
A more detailed tidal model, including eccentricity damping and planetary deformation, will be explored in future work.

\subsection{Dynamical Interaction of Multi-body Systems and Its Outcome}
\label{sec:dynamical_interact}

As the disc gas depletes, the timescale for eccentricity damping via gas drag becomes longer than the orbital crossing timescale (i.e., the time required for the closest planetary pair to destabilise and cross orbits). In such cases, orbital crossing occurs.

We adopt the semi-analytical model developed by \cite{Kimura+etal2025a,Kimura+2025b} to simulate the dynamical evolution of planetary systems via giant impacts and gravitational scattering, and to compute the resulting semi-major axes, eccentricities, and planetary masses.
In the following, we briefly summarise the numerical procedure. For details, see \cite{Kimura+etal2025a,Kimura+2025b}.

\begin{enumerate}
    \item \textbf{Identification of dynamically interacting planets} \\
    Orbital crossings or close encounters occur when eccentricity damping due to gas drag and dynamical friction from planetesimals becomes inefficient. To identify such planets, we first evaluate the orbital crossing timescale $\tau_{\rm cross}$ for all adjacent planetary triplets using the analytical formula from \cite{Petit+2020}, with an additional factor $K$ representing the density of three-body resonances, which depends on the number $N$ of nearby planets. See \cite{Kimura+2025b} for the evaluation of $K$ and $N$.

    Planets are considered dynamically interacting if they satisfy both:
    \begin{enumerate}
        \item $\tau_{\rm cross} < \tau_{\rm damp}$, where $\tau_{\rm damp}$ is the eccentricity damping timescale (Eq.~\eqref{eq:tau_dampTW}).
        \item $M_{\rm p} > M_{\rm pl,FZ}$, where
        \begin{equation}
            M_{\rm pl,FZ} := \int_{a_{\rm p}-\Delta a_{\rm FZ}/2}^{a_{\rm p}+\Delta a_{\rm FZ}/2}2\pi r \Sigma_{\rm s} \dd{r}
        \end{equation}
        is the total mass of planetesimals in the planet’s feeding zone, the width of which is given by $\Delta a_{\rm FZ} = b_{\rm FZ} r_{\rm H}$.
        \rev{
        This criterion follows \cite{Ida+Lin2010}. The eccentricity damping timescale due to dynamical friction scales inversely with the surrounding planetesimal surface density~\citep{Ida1990,Ohtsuki+2002}. Once the planet mass exceeds the local planetesimal mass, dynamical friction can no longer efficiently balance the mutual stirring between protoplanets~\citep[e.g.,][]{Goldreich+etal2004,Kokubo+Ida2012}.}
    \end{enumerate}
    Other planets are assumed to evolve independently.

    \item \textbf{Eccentricity evaluation at each timestep} \\
    The eccentricities of dynamically interacting planets are calculated using secular perturbation theory, considering all other planets in the system. Eccentricities of non-interacting planets (i.e., planets embedded in the disc gas or in the swarm of planetesimals) are randomly drawn from a Rayleigh distribution with a mean equal to their Hill eccentricity, $(M_{\rm p}/3M_*)^{1/3}$~\citep{Ida+Lin2010}. This initial condition has little impact on the subsequent dynamical evolution.

    \item \textbf{Timing of orbital crossing events} \\
    We re-evaluate $\tau_{\rm cross}$ using the updated eccentricities. Among the triplet with the shortest $\tau_{\rm cross}$, the pair $(i,j)$ with the smallest orbital separation undergoes an orbital crossing at
    \begin{equation}
        t_{\rm event} = t+\tau^*_{\rm cross}+\min(\tau_{{\rm scat},ij}, \tau_{{\rm col},ij}),
    \end{equation}
    where $\tau^*_{\rm cross}$ is the minimum crossing timescale, and $\tau_{{\rm scat},ij}$ and $\tau_{{\rm col},ij}$ are the scattering and collision timescales, respectively. We assume $a_i < a_j$.

    \item \textbf{Collision probability of the orbit-crossing pair} \\
    At $t = t_{\rm event}$, the eccentricities of the pair are set to $(e_{{\rm cross},i}, e_{{\rm cross},j})$, the minimum values required for orbital crossing. The outcome (giant collision or close scattering) is determined by using the ``collision probability'' $p_{\rm col}$ defined in ~\cite{Kimura+etal2025a}, which is a function of the ratio of $\tau_{{\rm scat},ij}$ to $\tau_{{\rm col},ij}$.

    \item \textbf{Eccentricity excitation via close encounters} \\
    We then compute the pre-event eccentricities $(e_{i0}, e_{j0})$ excited through repeated encounters. The relative eccentricity $e_{ij}$ is randomly drawn from a Rayleigh distribution with
    \begin{equation}
        \langle e^2_{ij}\rangle^{1/2} = e_{{\rm esc},ij} := \frac{\sqrt{2G(M_i+M_j)/(R_i+R_j)}}{\sqrt{GM_*/a_{ij}}},
    \end{equation}
    where $a_{ij} = \sqrt{a_i a_j}$. Individual eccentricities are assigned assuming energy equipartition.

    \item \textbf{Post-event orbital elements} \\
    In the case of a giant collision, the planets merge perfectly. The semi-major axis and eccentricity of the merged planet are computed using conservation of the centre of mass and the mass-weighted Laplace–Runge–Lenz vector.    
    If $e_{{\rm esc},ij} > 1$, we assume a collision occurs before significant eccentricity excitation, and set $(e_{i0}, e_{j0}) = (e_{{\rm cross},i}, e_{{\rm cross},j})$.

    In the case of close scattering, the eccentricities are set to $(e_{i0}, e_{j0})$, and the orbital separation is increased by the excited epicycle amplitudes. If either $e_i$ or $e_j$ exceeds unity, the planet is ejected, and the remaining planet’s orbital elements are updated using conservation of orbital energy.
\end{enumerate}

These steps are repeated until the integration time is reached or the number of planets is reduced to two. When only two planets remain, the orbital crossing timescale is no longer applicable. Instead, we assess system stability using the Jacobi energy in Hill coordinates~\citep{Nakazawa+Ida1988}.

\subsection{Initial Conditions and Parameters}
\label{sec:setting}

To initiate planetary population synthesis simulations, we performed random sampling of the initial stellar rotation rate $\Omega_{*,0}$, the disc gas mass $M_{\rm disc}$, the disc gas radius $r_{\rm disc}$, the disc metallicity [Fe/H], and the radius of the inner edge $r_{\rm in}$ of the disc. 
The distribution of these Monte Carlo variables are the same as KI22, except for $r_{\rm disc}$.
We then determine the initial masses and semi-major axes of planetary embryos in each disc, whose method is updated from KI22.
In the following, we briefly describe the procedure used to set these initial conditions.

\subsubsection{Initial Conditions for the Star and Protoplanetary Disc}
\label{sec:init_cond_disc}

The initial stellar rotation rate $\Omega_{*,0}$ is required to determine the evolutionary tracks of $L_{\rm X}$ and $L_{\rm EUV}$, based on the model of \cite{Johnstone+2021}. This model provides tabulated tracks corresponding to the $i$-th percentile of the observed distribution of stellar rotation rates. We randomly sample a percentile $i$ (denoted $i_{\Omega}$) and adopt the corresponding track to describe the time evolution of $L_{\rm X}$ and $L_{\rm EUV}$.

The initial properties of the protoplanetary disc are determined from recent observational studies. For the disc gas mass \rev{around a solar-mass star}, we adopt a log-normal distribution with mean \rev{$\log(\mu/M_\odot) = -1.49$} and standard deviation $\sigma = 0.35$, as derived by \citet{Emsenhuber+2021b} from the data of \cite{Tychoniec+2018}. The minimum and maximum disc masses are set to $4\times 10^{-3}M_\odot$ and $0.16M_\odot$, respectively, corresponding to the smallest and heaviest observed samples.
\rev{
The gas mass distribution is obtained by converting observed dust masses assuming a gas-to-dust ratio of 100. Because we independently sample the disc metallicity (see below), the resulting solid mass distribution differs from that inferred directly from \cite{Tychoniec+2018}. In addition, the observed sample is implicitly assumed to correspond to solar-mass stars. These assumptions introduce a conceptual inconsistency, also noted by \cite{Emsenhuber+etal2023a}. 
We nevertheless adopt the same prescription as \cite{Emsenhuber+2021b} to enable a direct comparison of planet population outcomes (see \S~\ref{sec:compare_NGPPS}). For stellar masses different from $1M_\odot$, the disc gas mass is scaled linearly with stellar mass~\citep{Andrews+2013}, acknowledging that the precise scaling remains uncertain~\citep[e.g.,][]{Ansdell+2016,Ansdell+2017,Testi+2022}.
}

The disc gas radius is then calculated using the empirical relation from \cite{Tobin+2020}:
\begin{equation}
    r_{\rm disc} = 70 \qty(\frac{M_{\rm disc}}{0.03M_\odot})^{0.25}~{\rm au}.
\end{equation}
The disc metallicity [Fe/H] follows a normal distribution with mean $\mu = -0.02$ and standard deviation $\sigma = 0.22$, based on \cite{Santos+2005}. The metallicity range is restricted to $-0.6 <$ [Fe/H] $< 0.5$. We use the same distribution regardless of the stellar mass.

The disc inner edge is assumed to be located at the corotation radius, where the Keplerian orbital period equals the stellar rotation period. The stellar rotation period is sampled from a log-normal distribution with $\log (\mu~{\rm [days]}) = 0.676$ and $\sigma = 0.306$, based on observations of young stellar objects by \cite{Venuti+2017}. The minimum value of $r_{\rm in}$ is set to the initial stellar radius $R_*$.
\rev{
We adopt this rotation-period distribution independent of stellar mass. Note that the sampled inner edge $r_{\rm in}$ is not required to be consistent with the independently sampled stellar rotation rate $\Omega_{*,0}$ used for the XUV evolution. This simplification allows the high-energy luminosity evolution and the disc inner boundary to vary independently.
}

\subsubsection{Initial Conditions for Planetary Embryos}
\label{sec:init_cond_planet}

The initial planetary mass is set to $0.01~M_\oplus$. The innermost planetary seed is placed at the innermost location of the solid (planetesimal) disc, which may differ from $r_{\rm in}$ due to the condition $\Sigma_{\rm s} = 0$ when $T_{\rm disc} > 0.99T_{\rm evap}$.

Subsequent planetary seeds are placed at intervals of the feeding zone $\Delta a_{\rm FZ} = b_{\rm FZ}r_{\rm H}$, where the mutual Hill radius $r_{\rm H}$ is calculated from the local asymptotic mass $M_{\rm max}$ at the final integration time $t_{\rm max}$. This mass is determined either by the local isolation mass~\citep{Kokubo+Ida1998,Ida+Lin2004} or by the mass that can be accreted via in-situ solid accretion within $t_{\rm max}$. See Appendix~\ref{sec:asymptotic_mass} for the explicit form and derivation of $M_{\rm max}$.
Planetary seeds are only placed in regions where $M_{\rm max} > 0.1M_\oplus$.
\rev{
This prescription is similar to that adopted by \cite{Ida+Lin2010} and is motivated by the oligarchic growth picture, in which embryos maintain separations of $\sim10$ mutual Hill radii until reaching the local isolation mass~\citep{Kokubo+Ida1998,Kokubo+Ida2002}. In regions where planets grow close to their asymptotic mass without significant migration (e.g. inside the snowline or in the outer disc), this approach provides a reasonable approximation.
However, just outside the snowline, planets undergo significant inward migration before reaching their asymptotic mass. In such cases, the initial separation (and thus the effective feeding zone) is overestimated, potentially leading to enhanced planetesimal accretion compared to a fully self-consistent treatment. Quantifying this effect and improving the embryo initialisation scheme are left for future work.
}

\subsubsection{Input Parameters}

The basic model parameters used throughout this study are listed in Table~\ref{tab:model_parameter}. These values remain fixed across all simulations.
\begin{table}
    \centering
    \caption{Model parameters}
    \label{tab:model_parameter}
    %\begin{tabularx}{\linewidth}{X|X|X|X}\hline
    \begin{tabular}{l|l|l|l} \hline 
    %{l|>{\centering}p{5cm}|l|l}\hline
         Quantity & Meaning & Value & Main Eqs. \\
         \hline \hline
         $t_{\rm max}$ & integration time & $5.0\times 10^9$~yr & \\ [0.2cm]
          $\alpha_{\rm acc}$ & 
         accretion viscosity $\alpha$ & $5.0\times 10^{-3}$ &
         Eq.~\eqref{eq:dSigma_dt_basic}\\ [0.2cm]
         $\alpha_{\rm mid}$ & mid-plane viscosity $\alpha$ &
         $5.0\times 10^{-4}$ & Eq.~\eqref{eq:Tvis_irr}\\ [0.2cm]
         $q_{\rm g}$ & power-law index of $\Sigma_{\rm g}$ & 0.9 & Eq.~\eqref{eq:Sigmag0} \\ [0.2cm]
         $q_{\rm s}$ & power-law index of $\Sigma_{\rm s}$ &
         1.5 & Eq.~\eqref{eq:Sigma_s} \\ [0.2cm]         
         $R_{\rm plt}$ & planetesimal radius &
         10~km &  \\ [0.2cm]
         $r_{\rm solid}$ & solid disc radius & $0.5r_{\rm disc}$ & Eq.~\eqref{eq:Sigma_s} \\ %[0.3cm]
         %
         % $\rho_{\rm plt}$ & planetesimal density & 
         % \begin{tabular}{l}
         %   $3.2~\si{g.cm^{-3}}$ ($r<r_{\rm ice}$) \\ $1.0~\si{g.cm^{-3}}$ ($r>r_{\rm ice}$) 
         % \end{tabular}
         % & Eq.~\eqref{eq:e_plt} \\ [0.3cm]
         %
        %  $C_{\rm rock}$ & 
        % \begin{tabular}{l}
        %   specific heat of solid core \\ for constant volume
        % \end{tabular}
        %  &  $\SI{1.2e7}{erg/(g.K)}$ & Eq.~\eqref{eq:Lcool} \\ 
         \hline
    \end{tabular}
\end{table}

\section{Example of Simulation Results}
\label{sec:example_simulation}

We present an example simulation illustrating the evolution of a protoplanetary disc and planets within our model framework. Unless otherwise stated, the initial conditions listed in Table~\ref{tab:ic_reference} are used throughout this section.

\begin{table}[t]
    \centering
    \caption{Initial conditions for the reference case.}
    \label{tab:ic_reference}
    \begin{tabular}{l|l|l}\hline
     Quantity & Meaning & Value \\ 
     \hline \hline
     $M_*$    & Stellar mass & $1M_\odot$  \\
     $M_{\rm disc}$  & Disc gas mass & $0.03M_\odot$ \\
     $[{\rm Fe/H}]$ & Disc metallicity & 0.0 \\
     $r_{\rm in}$ & Disc inner edge radius & 0.05~au \\
     $i_{\Omega}$ & Percentile index of $\Omega_{*,0}$ distribution & 50 \\ \hline          
    \end{tabular}
\end{table}

\subsection{Disc Evolution}

\begin{figure}
    \centering
    \includegraphics[width=\columnwidth]{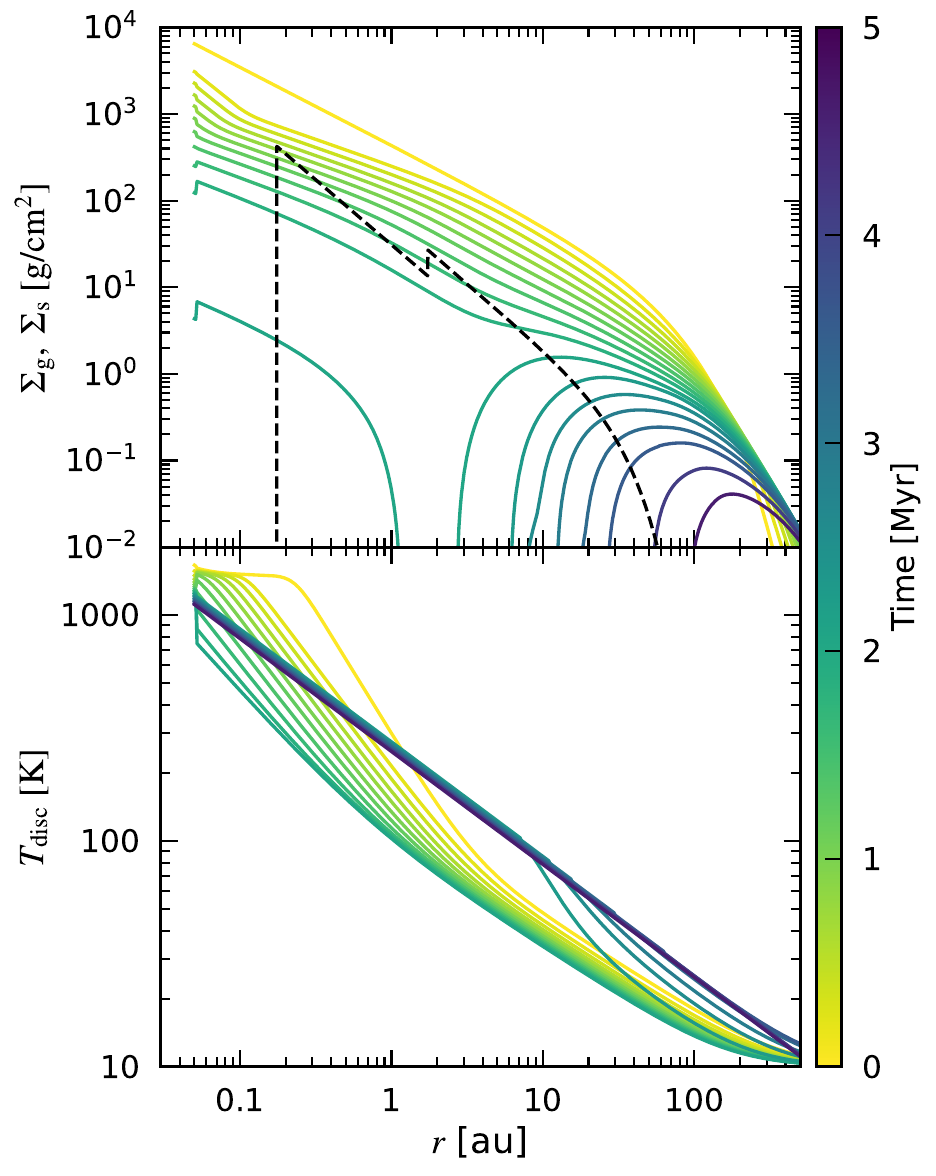}
    \caption{Time evolution of the disc gas surface density $\Sigma_{\rm g}$ (top) and mid-plane temperature $T_{\rm disc}$ (bottom), based on the initial conditions in Table~\ref{tab:ic_reference}. In the top panel, the initial solid surface density $\Sigma_{\rm s}$ is shown as a dashed line.}
    \label{fig:disk_gas}
\end{figure}

Figure~\ref{fig:disk_gas} shows the evolution of the disc gas surface density $\Sigma_{\rm g}$ and mid-plane temperature $T_{\rm disc}$. 
In this case, the disc lifetime, which is defined as the time when the vertical optical depth $\tau = \kappa \Sigma_{\rm g}$ drops below unity in regions with $T_{\rm disc} > 300$~K~\citep{Kimura+2016}, is $\sim$2.1~Myr, consistent with observational estimates~\citep[e.g.][]{Mamajek2009,Ansdell+2017}. Complete disc dissipation (i.e. $\tau < 1$ throughout the disc) occurs at 6.8~Myr.

The disc evolution follows the standard picture of viscous accretion and photoevaporation~\citep[e.g.][]{Clarke+2001,Alexander+2006a,Owen+2012,Suzuki+2016,Kunitomo+2020}. Initially, viscous accretion dominates until $\sim$1~Myr, after which a gap opens around 2~au due to strong X-ray photoevaporation. The inner disc then dissipates via accretion, while the outer disc is cleared by photoevaporation.

Regarding the temperature structure, viscous heating dominates the inner disc ($\lesssim 1$~au) in the early phase, while indirect stellar irradiation controls the temperature in the outer regions. After $\sim$1~Myr, irradiation becomes the dominant heating source throughout the disc. Following disc dispersal ($\gtrsim 2$~Myr), the equilibrium temperature $T_{\rm eq}$, calculated from the stellar bolometric luminosity, is plotted.

The top panel of Fig.~\ref{fig:disk_gas} also shows the initial solid surface density $\Sigma_{\rm s}$ (dashed line). In this setup, $\Sigma_{\rm s} = 0$ inside \rev{$\sim$0.2~au} due to sublimation at high temperatures. The iceline is located at $\sim$2~au, where $\Sigma_{\rm s}$ increases by a factor of two.

\subsection{Growth and Evolution of a \rev{Terrestrial Planet}}
\label{sec:one_planet}

\begin{figure}
    \centering
    \includegraphics[width=\linewidth]{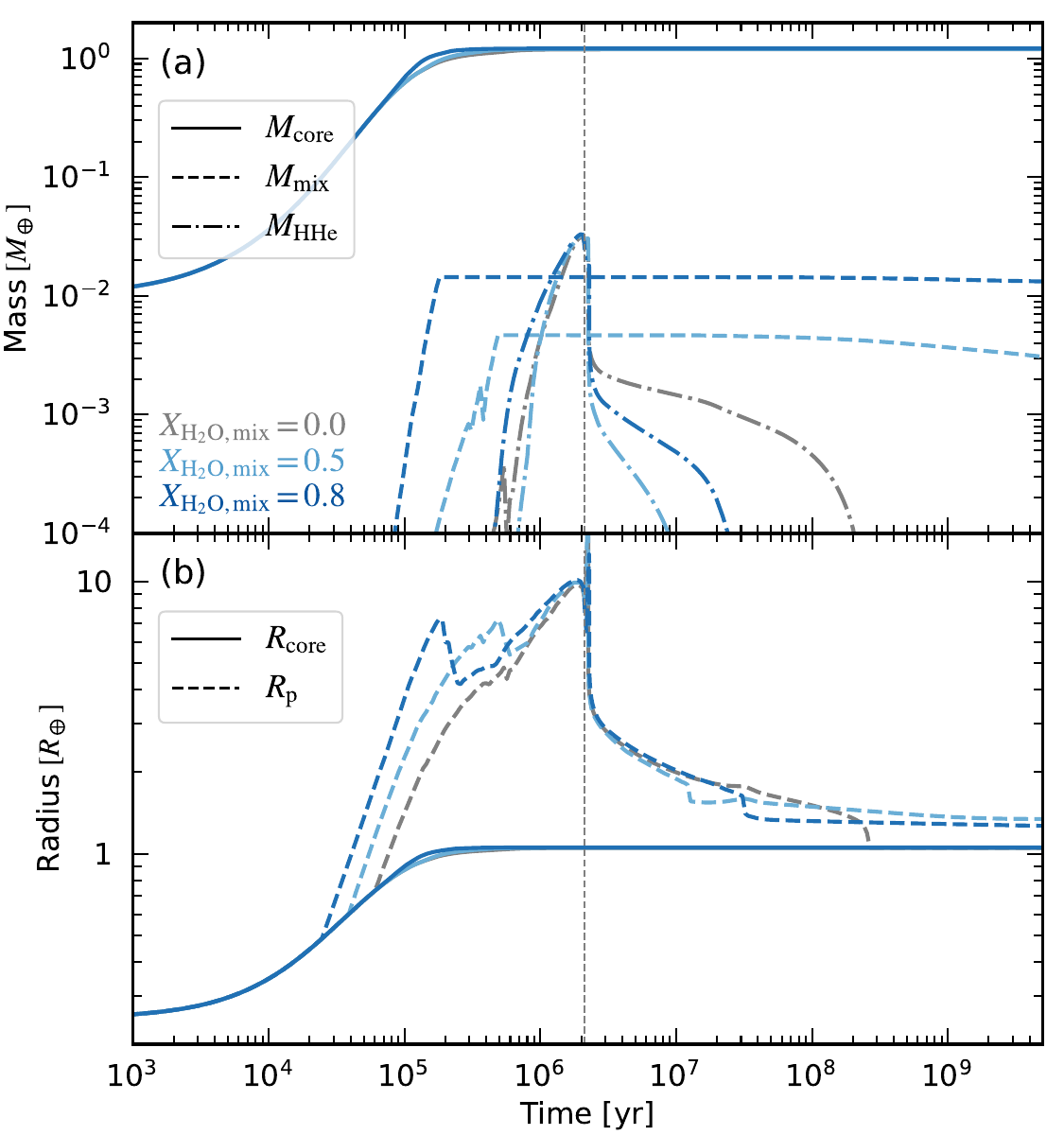}
    \caption{Evolution of the mass and radius of a planet located at 0.6~au in a disc with metallicity [Fe/H] = 0.2. All other parameters are as listed in Table~\ref{tab:ic_reference}. Results from three simulations with different water mass fractions in the vapour-mixed envelope, $X_{\rm H_2O,mix}$, are shown in different colours (grey: 0.0, light blue: 0.5, dark blue: 0.8). Panel (a) shows the mass of the solid core ($M_{\rm core}$; solid lines), vapour-mixed envelope ($M_{\rm mix}$; dashed lines), and H-He envelope ($M_{\rm HHe}$; dot-dashed lines). Panel (b) shows the solid core radius ($R_{\rm core}$; solid lines) and total planetary radius ($R_{\rm p}$; dashed lines), defined at the 10~mbar pressure level. The vertical dashed line indicates the disc dispersal time ($\sim$2.1~Myr).}
    \label{fig:one_planet}
\end{figure}

Figure~\ref{fig:one_planet} illustrates the growth and evolution of a planet placed at 0.6~au with an initial mass of $0.01M_\oplus$. Orbital migration is not included. The disc metallicity is set to [Fe/H] = 0.2 to allow sufficient in-situ growth for envelope accumulation. 
Other disc parameters follow Table~\ref{tab:ic_reference}. We compare three cases with different water mass fractions in the vapour-mixed envelope: $X_{\rm H_2O,mix} = 0.0$, 0.5, and 0.8.

The planetary core reaches its isolation mass of $\sim 1M_\oplus$ at $\sim$0.2~Myr. Envelope accretion begins after planetesimal accretion ceases.
In the case with $X_{\rm H_2O,mix} = 0.0$, no vapour-mixed layer forms, and only an H-He envelope is accreted by definition. At disc dispersal, the H-He envelope mass reaches $\sim$2\% of the total planetary mass. Due to the inflated planetary radius ($\sim 8R_\oplus$), the escape parameter is low, resulting in rapid photoevaporation~\citep{Kubyshkina+2018a}. The radius shrinks to $\sim$2–3$R_\oplus$ within $\sim$1~Myr, and the envelope is completely lost by $\sim$0.3~Gyr.

In the enriched cases ($X_{\rm H_2O,mix} = 0.5$ and 0.8), the vapour-mixed envelope forms before isolation. The envelope mass $M_{\rm mix}$ is larger for $X_{\rm H_2O,mix} = 0.8$ due to the higher mean molecular weight and lower adiabat, which alters the envelope structure~\citep{Kimura+Ikoma2020}. 
%Isolation occurs slightly earlier for higher $X_{\rm H_2O,mix}$, as the denser envelope enhances the planetesimal capture radius \citep{Inaba+Ikoma2003}. Although the difference is minor in the case (0.5~au) presented in Fig.~\ref{fig:one_planet}, it becomes more significant for planets at larger orbital distances.
The H-He envelope mass on top of the vapour-mixed envelope is similar across all cases and is rapidly lost after disc dispersal. However, the vapour-mixed envelope remains, as its high mean molecular weight leads to a smaller geometric cross-section and lower escape rates~\citep{Yoshida+2022,Yoshida+Gaidos2025}. Consequently, the final planetary radius is larger in the enriched cases.

These results demonstrate that water enrichment in the envelope significantly influences planetary growth, envelope evolution, and final properties such as radius, mass, and composition.

\rev{
\subsection{Growth and Evolution of a Giant Planet}
}
\label{sec:growth_giant}

\begin{figure}
    \centering
    \includegraphics[width=\linewidth]{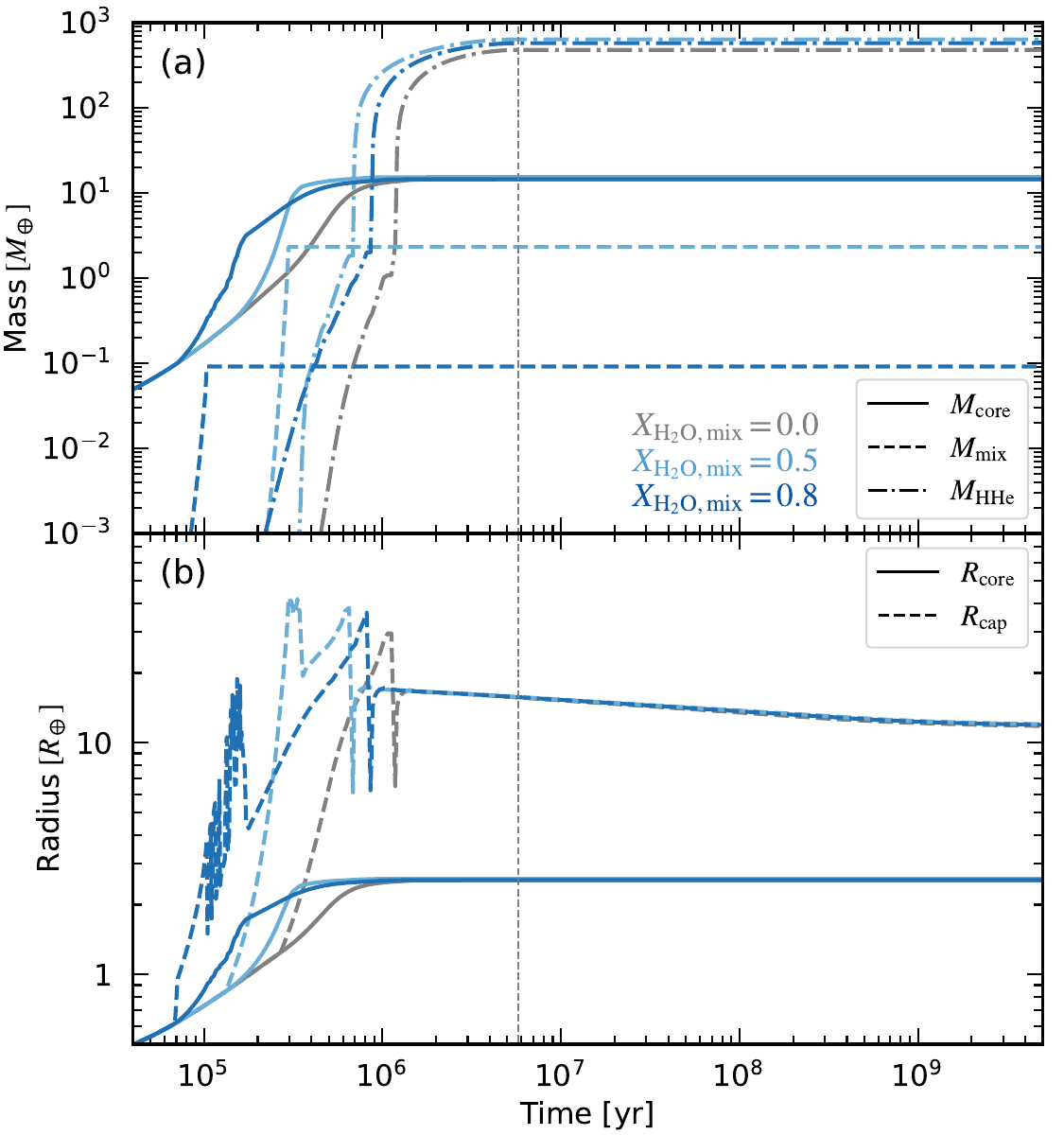}
    \caption{
    \rev{
    Same as Fig.~\ref{fig:one_planet}, but for a planet located at 2.5~au in a disc with $M_{\rm disc}=0.06M_\odot$ and $[{\rm Fe/H}] = 0.2$. Other parameters follow Table~\ref{tab:ic_reference}. 
    Note that the early fluctuations of $R_{\rm cap}$ in the $X_{\rm H_2O,mix}=0.8$ case are caused by the accumulation of a very small H–He mass ($\sim10^{-8}M_\oplus$) above the vapour-mixed layer.
    }
    }
    \label{fig:oneplanet_giant}
\end{figure}

\rev{
Figure~\ref{fig:oneplanet_giant} presents an example of in-situ growth leading to a giant planet. The initial embryo is placed at 2.5~au, and orbital migration is not included. The disc mass and metallicity are set to $M_{\rm disc}=0.06M_\odot$ and $[{\rm Fe/H}]=0.2$ to allow efficient core growth. Other disc parameters follow Table~\ref{tab:ic_reference}. 
We compare three cases with different vapour-mixed water fractions, $X_{\rm H_2O,mix}=0.0$, 0.5, and 0.8.
}

\rev{
Although the final planet mass and radius are similar among the three cases, the core growth timescale and the onset of runaway gas accretion depend on $X_{\rm H_2O,mix}$. Runaway accretion (Phase~III) begins at $\sim1$~Myr, marked by the rapid increase in $M_{\rm HHe}$ in Fig.~\ref{fig:oneplanet_giant}(a).
During Phase~III, the internal structure is no longer solved explicitly; the planetesimal capture radius is taken from the precomputed giant-planet table (\S~\ref{sec:evolution_thick_envelope}). This causes the discontinuity in $R_{\rm cap}$ seen in Fig.~\ref{fig:oneplanet_giant}(b). The impact on solid accretion is minor, as most planetesimals are ejected rather than accreted in this phase.
}

\rev{
The dependence of core growth on $X_{\rm H_2O,mix}$ arises because a more enriched envelope is more massive and denser, which increases the planetesimal capture radius (Fig.~\ref{fig:oneplanet_giant}(b)). Consequently, larger $X_{\rm H_2O,mix}$ generally accelerates core growth and advances the onset of runaway accretion, consistent with the mechanism reported by \cite{Venturini+2016}.
}

\rev{
However, the effect does not increase monotonically with $X_{\rm H_2O,mix}$. In our simulations, runaway accretion begins earlier for $X_{\rm H_2O,mix}=0.5$ than for $0.8$. 
At very high enrichment, the critical core mass is substantially reduced~\citep{Hori+Ikoma2011,Venturini+2015}, allowing the transition from Phase I to II to occur during the planetesimal accretion stage.
The vapour-mixed envelope then contracts rapidly and H–He accretion starts. 
Once the H–He layer controls the capture radius, solid accretion becomes less efficient. This is reflected in the sharp decrease of $R_{\rm cap}$ at $\sim0.2$~Myr in the $X_{\rm H_2O,mix}=0.8$ case, after which core growth slows relative to the $X_{\rm H_2O,mix}=0.5$ case.
These results suggest that giant planet formation is most efficient at intermediate values of $X_{\rm H_2O,mix}$, a trend further discussed in \S~\ref{sec:depend_X}.
}

\begin{figure}
    \centering
    \includegraphics[width=\columnwidth]{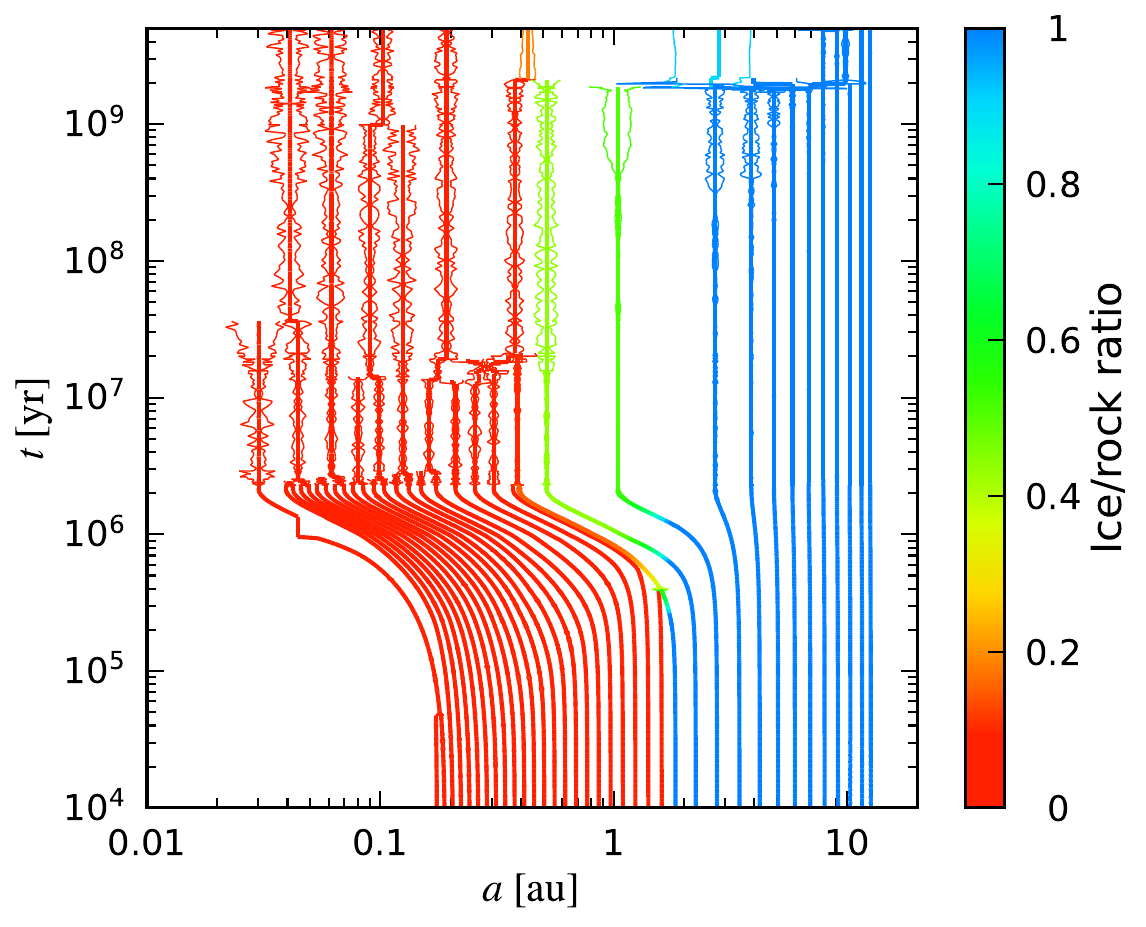}
    \caption{Orbital and collisional evolution of planets in a planetary system simulated with the initial conditions in Table~\ref{tab:ic_reference}. Thick lines indicate the semi-major axes of planets, while thin lines show their periapsis and apoapsis. Colours represent the ice-to-rock ratio of the planetary cores.}
    \label{fig:track_ecc}
\end{figure}

\begin{figure}
    \centering
    \includegraphics[width=\columnwidth]{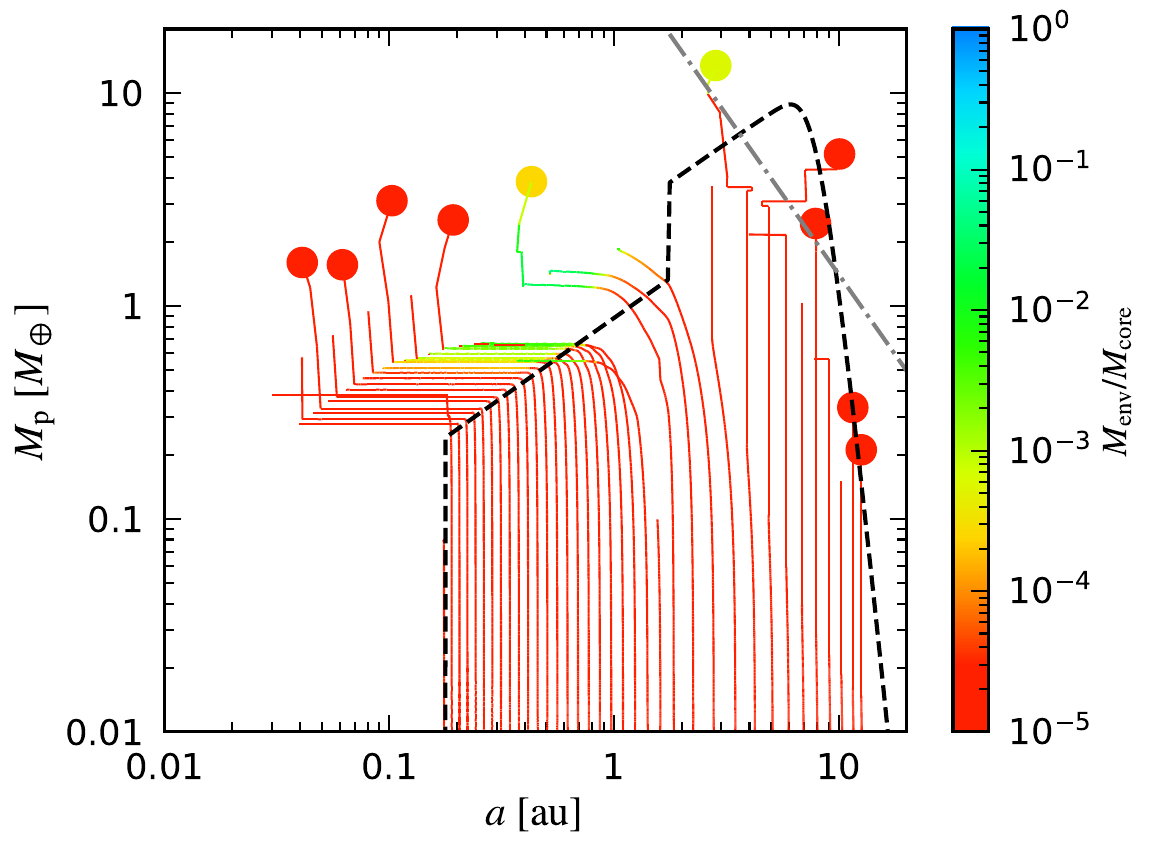}
    \caption{Evolution of the mass and semi-major axis of planets in the same system as in Fig.~\ref{fig:track_ecc}. Circles indicate the final states of planets; lines show their evolutionary tracks. Tracks are also shown for planets that experienced collisions. Colours represent the envelope mass fraction at each timestep. The dashed line shows the local asymptotic mass $M_{\rm asym}$ (Eq.~\eqref{eq:M_asym}), and the dot-dashed line shows the mass threshold for planetesimal ejection $M_{\rm esc}$ (Eq.~\eqref{eq:M_esc}).}
    \label{fig:formation_track}
\end{figure}

\begin{figure*}[t]
    \centering
    \includegraphics[width=1.7\columnwidth]{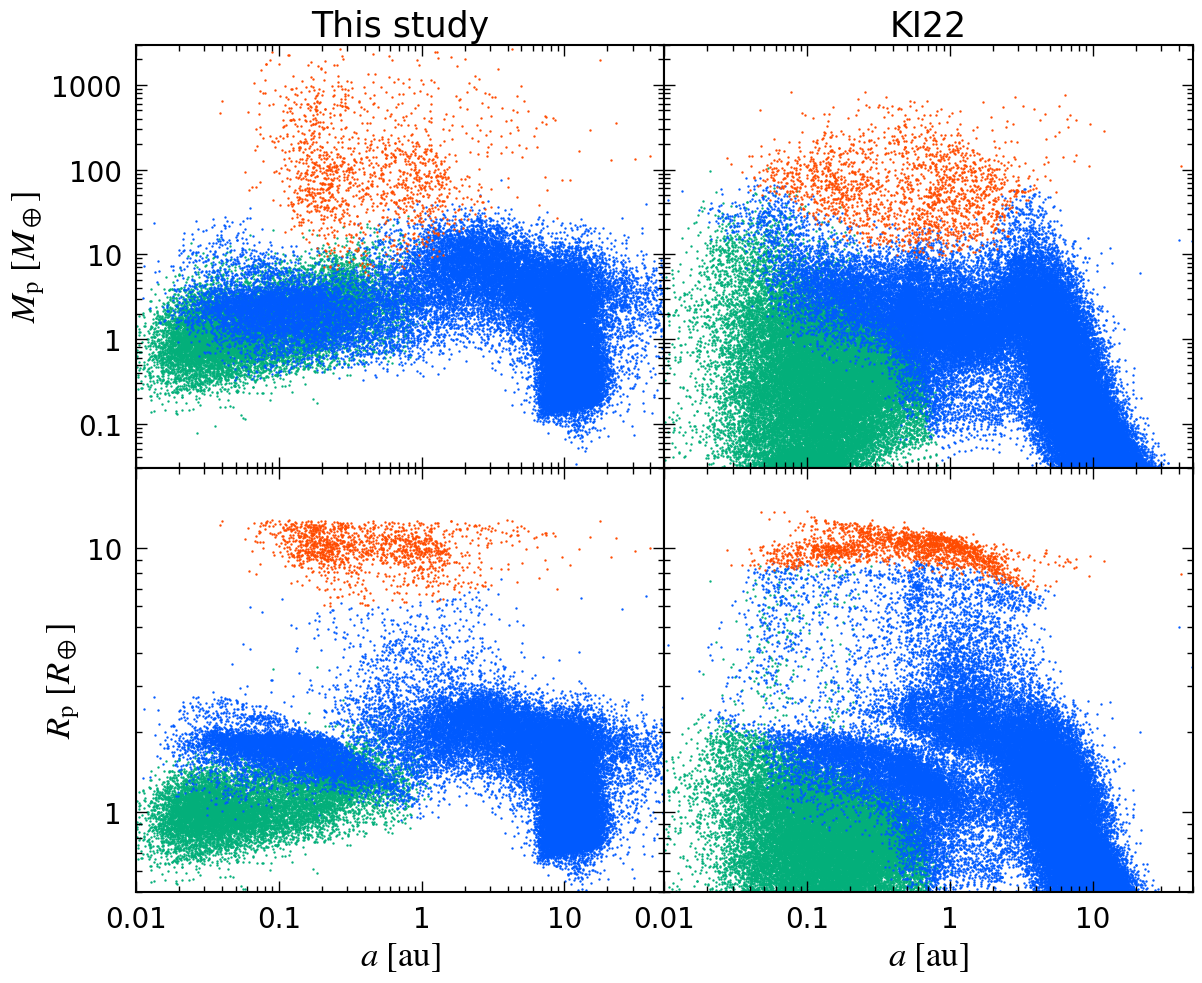}
    \caption{Final distributions of planetary mass (top row) and radius (bottom row) versus semi-major axis for 5000 systems. Left: results from this study; right: results from \cite{Kimura+Ikoma2022}. Both cases assume $M_* = 1M_\odot$ and $X_{\rm H_2O,mix} = 0.0$. Note that the integration time is 5~Gyr in this study and 1~Gyr in the previous study. Red points indicate planets with envelope mass fraction $M_{\rm env}/M_{\rm core} > 1$. Among the rest, blue and green points represent planets with and without ice in the core, respectively.}
    \label{fig:Ma_Ra}
\end{figure*}

\subsection{Evolution of a Planetary System}
We simulate the dynamical evolution of an entire planetary system from the initial conditions in Table~\ref{tab:ic_reference}, focussing on the impact of updated treatments for planet–planet dynamical interactions. In this example, we set $X_{\rm H_2O,mix} = 0.0$ for all planets.

Figure~\ref{fig:track_ecc} shows the evolution of planetary semi-major axes and eccentricities. During the disc phase ($\lesssim$~2.1~Myr), planets grow and migrate inward, forming a resonant chain from the disc inner edge ($\sim$0.05~au) to $\sim$0.4~au. While some planets collide without being trapped in resonance, most undergo resonance trapping due to convergent migration. Planetary masses at disc dispersal are determined primarily by the local isolation mass or by the point at which migration removes the planet from its initial feeding zone.
Note that, although a migrating planet can accrete planetesimals along its migration path, this additional accretion is generally minor because most planetesimals in the inner disc have already been accreted by planets that formed in situ.

After disc dispersal, close-in planets become dynamically unstable, leading to frequent collisions. These features are consistent with previous $N$-body-based formation models~\citep[e.g.][]{Emsenhuber+2021a}.

In contrast, planets beyond $\sim$1~au continue to grow via planetesimal accretion after disc dispersal. Initially, they remain dynamically stable due to the presence of surrounding planetesimals. However, by $\sim$1~Gyr, some planets have cleared their feeding zones and begin to interact gravitationally, triggering late-stage instabilities. This is evident at $\sim$2~Gyr in Fig.~\ref{fig:track_ecc}, where distant planets undergo scattering and collisions, altering their orbits and masses.

Figure~\ref{fig:formation_track} shows the final masses and semi-major axes of the planets, along with their formation tracks. The local asymptotic mass $M_{\rm asym}$, determined by either isolation or accretion timescale limits, is also plotted with black dashed lines. Inner planets typically reach their isolation mass before migrating inward. In contrast, outer planets, especially those beyond the iceline, begin migrating before reaching isolation. Their masses at the migration phase are set by the balance between core growth and migration timescales~\citep[see also][]{Emsenhuber+2023}.
Despite the large mass difference between the inner and outer planets in the disc phase, the inner planets significantly grow via giant collisions after the disc dispersal, leading to a small intra-system mass deviation at the final state.

Envelope formation, indicated by colour in Fig.~\ref{fig:formation_track}, occurs mainly during the migration phase, after planetesimal accretion has ceased. Inner planets, being sub-Earth mass at this stage, accrete only thin envelopes, which are quickly lost after disc dispersal. In contrast, planets formed beyond the iceline grow larger and retain more substantial envelopes, some of which survive until the end of the simulation. Planets beyond $\sim$5~au remain envelope-free due to insufficient growth during the disc lifetime.

%In this simulation, the final system consists of four rocky planets (Earth- to super-Earth-mass) without envelopes and one icy super-Earth-mass planet with an envelope, all located within $\sim$1~au.

\section{Synthetic Population}
\label{sec:Monte_Carlo}

We perform Monte Carlo simulations by randomly sampling initial conditions as described in \S~\ref{sec:init_cond_disc}, to generate synthetic planetary populations. A total of 12 sets of simulations are conducted, varying stellar mass $M_*$ (0.1, 0.3, 0.5, and 1~$M_\odot$) and water mass fraction in the vapour-mixed envelope $X_{\rm H_2O,mix}$ (0.0, 0.5, and 0.8). For each parameter set, 5000 planetary systems are simulated.

We examine how the modules updated in this study affect the final distributions of simulated planets compared to those from our previous study~\citep{Kimura+Ikoma2022}, and how envelope enrichment and stellar mass influence these distributions.

\subsection{Nominal Case}

Figure~\ref{fig:Ma_Ra} compares the final distributions of planetary mass (\textit{top}) and radius (\textit{bottom}) versus semi-major axis between this study (\textit{left}) and our previous study (KI22; \textit{right}), for the nominal case with $M_* = 1M_\odot$ and $X_{\rm H_2O,mix} = 0.0$. Note that the integration time is extended to 5~Gyr in this study, compared to 1~Gyr in KI22, to capture dynamical instabilities that typically occur at late stages of evolution ($\gtrsim$1~Gyr), as illustrated in Fig.~\ref{fig:track_ecc}.

We find several notable differences. Our updated model produces fewer low-mass planets (sub-Earth masses), while KI22 shows a significant population of such planets. Conversely, KI22 yields many close-in planets with $M_{\rm p} > 10M_\oplus$, which are rare in this study's results. Namely, the updated PPS model brings about a narrower mass distribution for super-Earths and smaller planets.

For giant planets ($M_{\rm env}/M_{\rm core} > 1$), both models show that most of the giant planets have sub-Jovian masses, but this study's model also produces a substantial number of super-Jovian planets, which are rare in KI22. Additionally, this study's model shows more icy planets at $\lesssim 0.1$~au, indicating more efficient inward migration. Those migrating icy planets cause many planets growing in situ to be engulfed by their host stars due to dynamical interactions or tidal migration, which was rarely found in KI22.

These differences stem from updates to the post-disc dynamical evolution and disc gas evolution. The reduced abundance of close-in sub-Earth-mass planets in this study results from more frequent giant impacts after disc dispersal, as predicted by the semi-analytical model of \cite{Kimura+etal2025a}. In contrast, KI22 used the model of \cite{Ida+Lin2010}, which predicts fewer collisions.
%Figure~\ref{fig:Ma_snp} shows the time evolution of the mass–semi-major axis distribution. Many sub-Earth-mass planets form by disc dispersal, and frequent collisions lead to rapid growth to Earth-like or super-Earth masses within $\sim$0.1~Gyr, 
Indeed, our updated model results show better agreement with $N$-body-based population synthesis results in terms of the mass distribution of such low mass planets~\citep{Emsenhuber+2021b,Emsenhuber+2023}.

\begin{figure}
    \centering
    \includegraphics[width=0.8\linewidth]{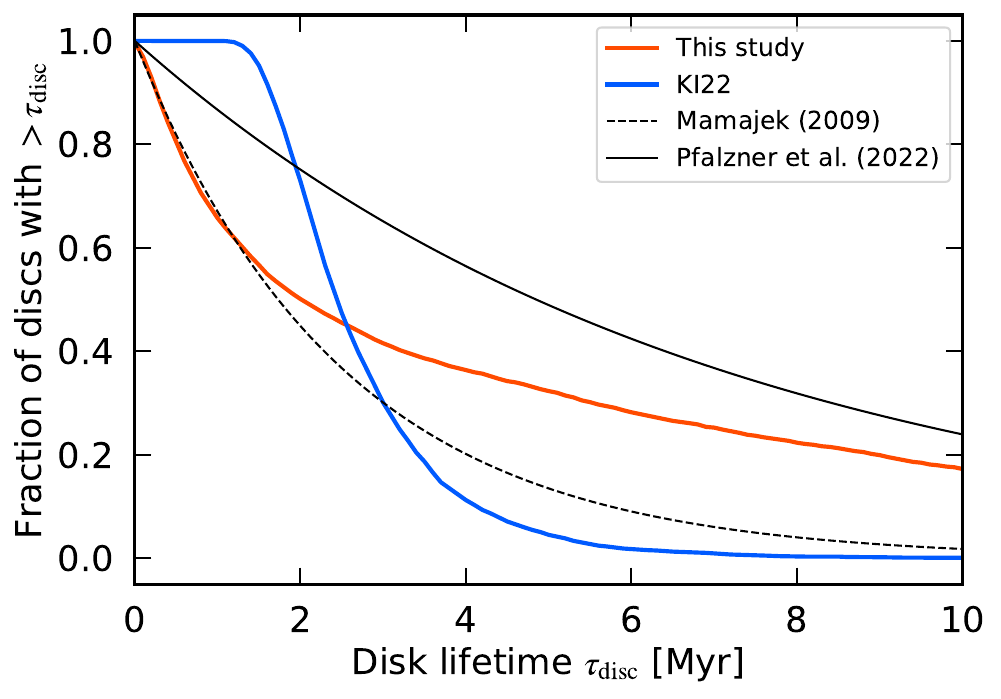}
    \caption{Distribution of disc lifetimes in this study's model (red) and in \cite{Kimura+Ikoma2022} (blue), for the nominal case with $M_* = 1M_\odot$ and $X_{\rm H_2O,mix} = 0.0$. Observational fits from \cite{Mamajek2009} (dashed black line) and \cite{Pfalzner+2022} (solid black line) are also shown.}
    \label{fig:disk_lifetime}
\end{figure}

Differences in giant planet masses and close-in planet distributions also arise from our updated disc photoevaporation model, which affects disc lifetimes. Figure~\ref{fig:disk_lifetime} compares disc lifetime distributions. KI22 shows a narrow distribution centred around 2–3~Myr, lacking both short-lived ($\lesssim$1~Myr) and long-lived ($\gtrsim$5~Myr) discs. By contrast, this study's model yields a broader distribution, including discs with $\tau_{\rm disc} > 10$~Myr.

In KI22, disc lifetimes were tuned via external photoevaporation rates to match observations~\citep{Mamajek2009,Ansdell+2017}. In contrast, our model determines disc lifetimes from stellar X-ray luminosity $L_{\rm X}$ and initial disc mass $M_{\rm disc}$, without tuning. While our lifetime distribution does not match the rapid exponential decay suggested by some studies~\citep[e.g.][]{Mamajek2009,Ansdell+2017,Richert+2018}, recent observations of nearby low-mass stars suggest longer decay timescales of $\sim$7~Myr~\citep{Pfalzner+2022}, consistent with our results.
Thus, the broader disc lifetime distribution in this study is observationally plausible and leads to more diverse planetary outcomes.

\rev{
The more efficient migration in the updated model also produces a shallow valley of icy planets at $\sim0.5$--1~au in the $R_{\rm p}$--$a$ distribution. While a similar feature is present in KI22, there it is primarily associated with the presence or absence of envelopes. In the present model, an additional contribution arises from a mass contrast between inner ($\lesssim1$~au) and outer ($\gtrsim1$~au) icy populations, which is evident in the $M_{\rm p}$--$a$ distribution.
The inner icy planets typically migrate inward before reaching their local isolation mass, whereas outer planets grow largely in situ after disc dispersal and approach isolation mass. This divergence in growth histories leads to a systematic mass difference between the two populations, which in turn shapes the radius valley. The distinct evolutionary tracks are also illustrated in Fig.~\ref{fig:formation_track}.
}

\subsection{Dependence on $X_{\rm H_2O,mix}$}
\label{sec:depend_X}
\begin{figure*}
    \centering
    \includegraphics[width=2\columnwidth]{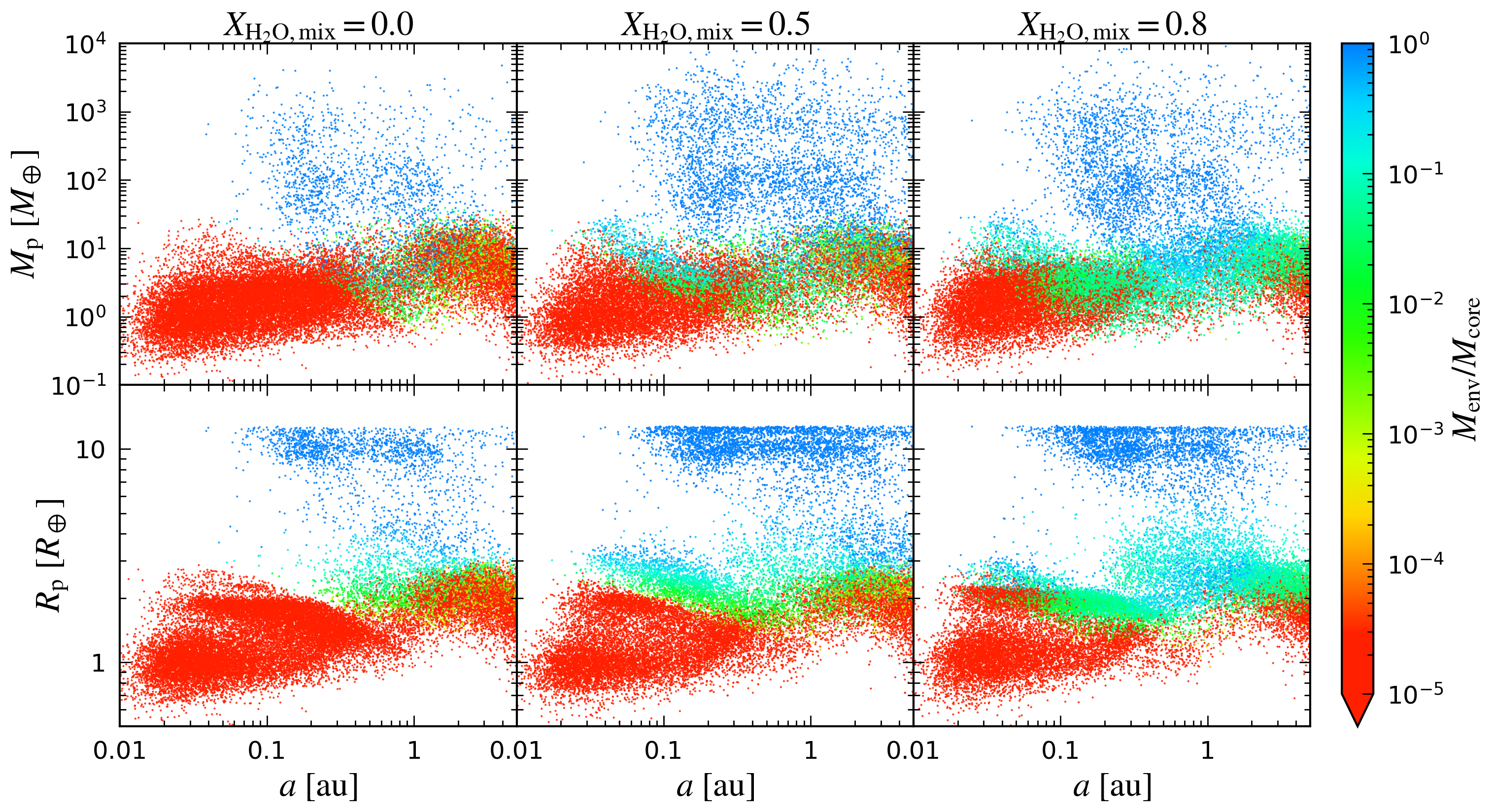}
    \caption{Final distributions of simulated planets for $X_{\rm H_2O,mix} = 0.0$ (left), 0.5 (centre), and 0.8 (right), with stellar mass fixed at $1M_\odot$. Colours indicate envelope mass fraction.}
    \label{fig:Ma_Ra_xH2O}
\end{figure*}

Figure~\ref{fig:Ma_Ra_xH2O} shows the final distributions of simulated planets for different values of $X_{\rm H_2O,mix}$, with stellar mass fixed at $1M_\odot$. The envelope enrichment significantly affects the radius and envelope mass of terrestrial and super-Earth planets, as well as the formation frequency of giant planets.

First, an increasing number of planets with masses ranging from several to several tens of Earth masses (and radii of 2-3 Earth radii) are found to form within 0.1~au in the cases of enriched envelopes (bluish circles).  
As discussed in \S~\ref{sec:one_planet}, planets with Earth-like to super-Earth masses can retain thick envelopes in the enriched cases ($X_{\rm H_2O,mix} = 0.5$ and 0.8), which survive photoevaporation. Consequently, many close-in super-Earths retain envelopes with mass fractions of $\sim$1–10\%, leading to inflated radii. %For example, in the case with $X_{\rm H_2O,mix} = 0.5$, a distinct population of planets with radii $\sim$2–3~$R_\oplus$ emerges, which is absent in the non-enriched case. In this setting, 
In the case of $X_{\rm H_2O,mix} = 0.5$, three clusters of close-in planets are observed:
rocky planets without envelopes in $\sim 1R_\oplus$,
icy planets with steam atmospheres in $\sim 2R_\oplus$,
planets with vapour-mixed envelopes at $\sim 2$–3~$R_\oplus$.
Similar populations appear for $X_{\rm H_2O,mix} = 0.8$, although the steam atmosphere and vapour-mixed envelope populations overlap due to the thinner vapour-mixed layer caused by the higher mean molecular weight.

Envelope enrichment also influences the formation frequency of giant planets. The occurrence rate of giants ($M_{\rm env}/M_{\rm core} > 1$) is approximately three times higher in the enriched cases than in the non-enriched case. This is because the enriched, \rev{massive} envelope enhances the planetesimal capture radius, accelerating core growth, \rev{as already found in \cite{Venturini+2016}.} 
Planets beyond the iceline, in particular, can reach the runaway gas accretion phase before migrating inward.

However, the effect does not increase simply with $X_{\rm H_2O,mix}$. The formation rate of giant planets for $X_{\rm H_2O,mix} = 0.8$ is slightly lower than for 0.5. 
\rev{As discussed in \S~\ref{sec:growth_giant}, extremely high enrichment can trigger early H–He accumulation during the planetesimal accretion phase, reducing the effective capture radius and slowing subsequent core growth.} 
%In highly enriched cases, the critical core mass decreases significantly~\citep{Hori+Ikoma2011,Venturini+2015}, allowing planets to reach this threshold during the planetesimal accretion phase. The vapour-mixed envelope then contracts rapidly, and H-He accretion begins. Since the planetesimal capture radius is now governed by the H-He envelope, the accretion rate slows and core growth becomes less efficient.

These results highlight the importance of envelope composition in shaping both low-mass and giant planet populations. Detailed modelling of envelope formation during the planet formation stage is essential for interpreting observed planetary distributions.

\begin{figure*}
    \centering
    \includegraphics[width=1.8\columnwidth]{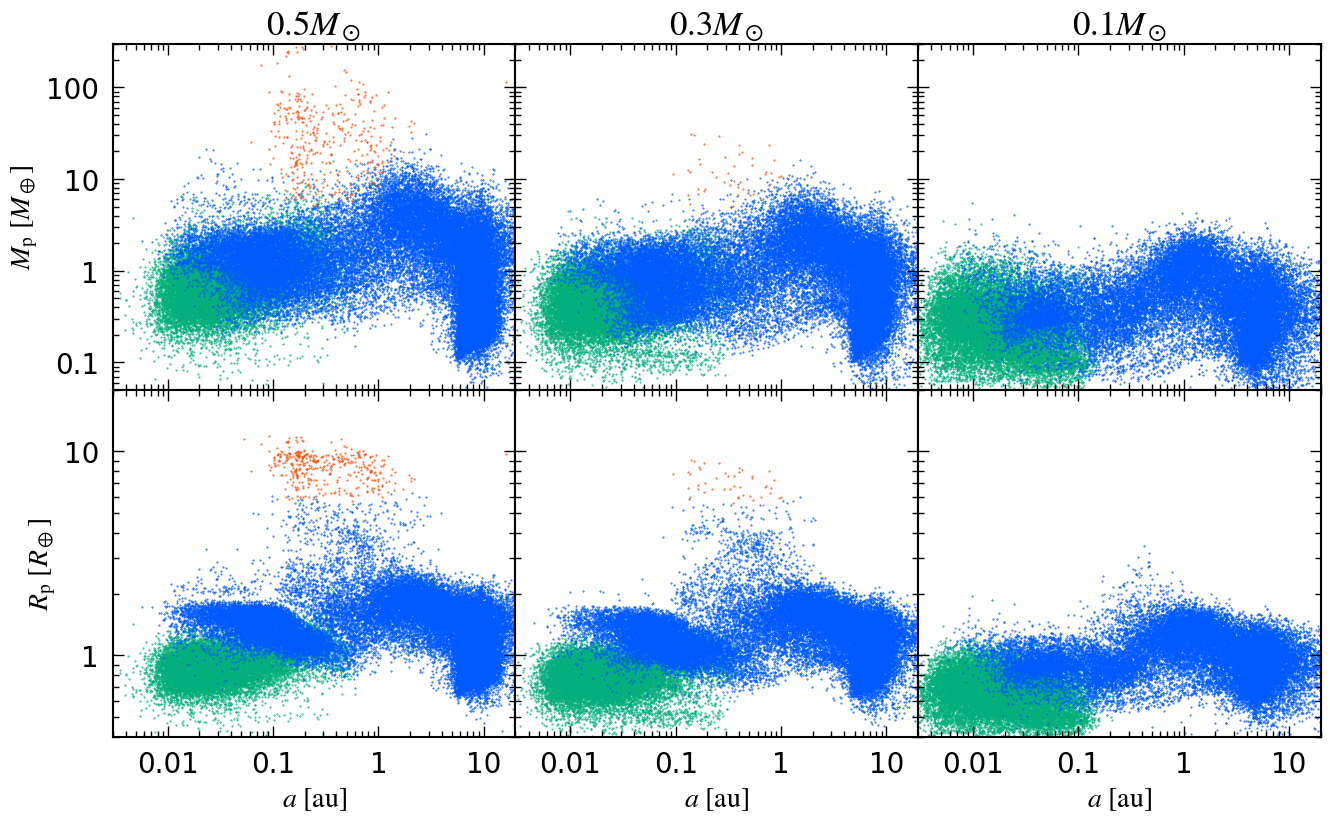}
    \caption{Same as Fig.~\ref{fig:Ma_Ra}, but for different stellar masses: $0.5M_\odot$ (left), $0.3M_\odot$ (centre), and $0.1M_\odot$ (right). All cases assume $X_{\rm H_2O,mix} = 0.0$. Red points indicate planets with envelope mass fraction $M_{\rm env}/M_{\rm core} > 1$. Among the rest, blue and green points represent planets with and without ice in the core, respectively.}
    \label{fig:Ma_Ra_Mstar}
\end{figure*}

% \begin{figure}
%     \centering
%     \includegraphics[width=0.8\columnwidth]{hist_water_nominal_occurrence.pdf}
%     \caption{Occurrence rate distribution of water mass fraction in simulated planets with masses of 0.3–3~$M_\oplus$ located in the habitable zone around stars of $0.1M_\odot$ (top), $0.3M_\odot$ (middle), and $0.5M_\odot$ (bottom). For each stellar mass, cases with $X_{\rm H_2O,mix} = 0.0$, 0.5, and 0.8 are shown in red, blue, and green, respectively. See text for the definition of the habitable zone.}
%     \label{fig:hist_Mw}
% \end{figure}

\subsection{Dependence on Stellar Mass}

Figure~\ref{fig:Ma_Ra_Mstar} shows the final distributions of simulated planets around M dwarfs with stellar masses of 0.1, 0.3, and 0.5~$M_\odot$, assuming $X_{\rm H_2O,mix} = 0.0$.  
Overall, both the typical planetary mass and the occurrence of giant planets decrease with decreasing stellar mass, consistent with recent population synthesis studies~\citep{Burn+2021}.  
In particular, no giant planets with masses $\gtrsim 100M_\oplus$ are produced around stars with $\lesssim 0.3M_\odot$.

The abundance of close-in planets that retain primordial envelopes (i.e. $\gtrsim 2R_\oplus$) is also markedly reduced around M dwarfs.  
This occurs because planets orbiting lower-mass stars generally cannot grow sufficiently massive through planetesimal accretion within the disc lifetime, resulting in less envelope accretion.  
Moreover, such planets experience stronger XUV irradiation, leading to more efficient envelope escape.  
Consequently, close-in planets around M dwarfs rarely retain enough gas to reach sub-Neptune sizes.

\section{Discussion}
\label{sec:discussion}

\subsection{Comparison with NGPPS}
\label{sec:compare_NGPPS}
\begin{figure*}
    \centering
    \includegraphics[width=2\columnwidth]{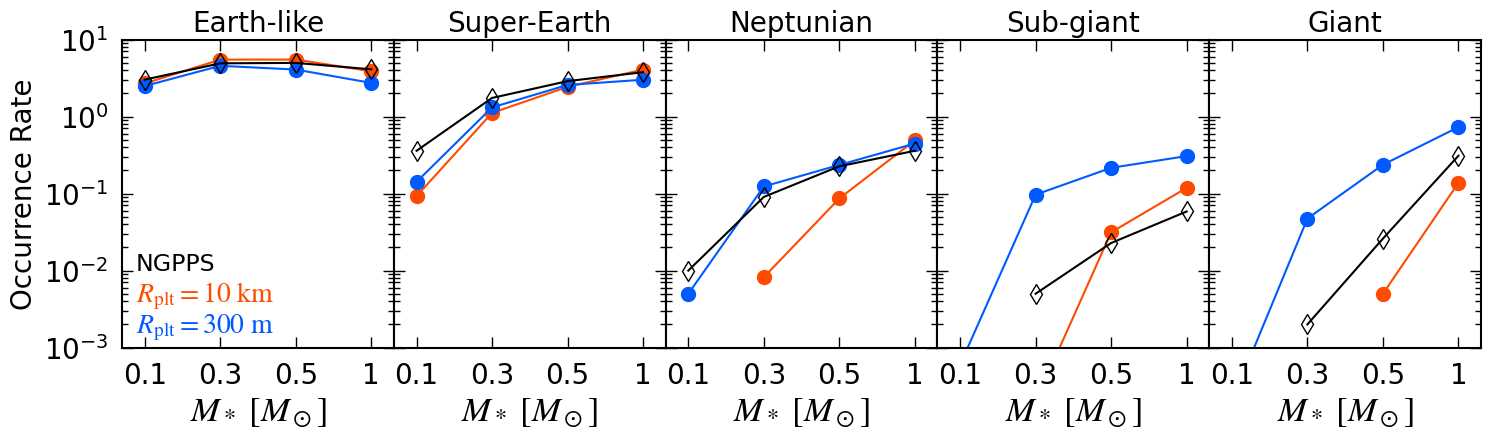}
    \caption{Comparison of occurrence rates by planet type between our simulation results (red and blue) and those from NGPPS (black), taken from \cite{Burn+2021}. Red: our nominal case with planetesimal radius $R_{\rm plt} = 10$~km; blue: case with $R_{\rm plt} = 300$~m, matching NGPPS. All cases assume $X_{\rm H_2O,mix} = 0.0$. See text for definitions of the five planet categories.}
    \label{fig:statics}
\end{figure*}

We briefly compare our simulation results with those from the New Generation Planet Population Synthesis (NGPPS)~\citep{Emsenhuber+2021a,Emsenhuber+2021b,Burn+2021}, which is the first population synthesis framework to incorporate detailed models of envelope formation and thermal evolution and planet–planet gravitational interactions.

Our model presented in this paper shares several features with NGPPS, including prescriptions for solid core growth, internal structure, and envelope thermal evolution (although NGPPS does not include water enrichment in the envelope), as well as similar initial conditions for Monte Carlo simulations. 
However, key differences exist in the treatment of disc structure and evolution, runaway gas accretion, envelope escape, orbital migration, and dynamical interactions (handled via direct $N$-body integration in NGPPS). Additionally, the assumed disc viscosity $\alpha$ and planetesimal radius differ. See \cite{Burn+Mordasini2025} for a comprehensive review of NGPPS and comparisons with other models, including KI22.

To assess how these differences affect the resulting planet populations, we focus on planetary occurrence rates. Following \cite{Burn+2021}, we classify simulated planets into five categories:
\begin{itemize}
    \item Earth-like: 0.5–2~$M_\oplus$
    \item Super-Earths: 2–10~$M_\oplus$
    \item Neptunians: 10–30~$M_\oplus$
    \item Sub-giants: 30–100~$M_\oplus$
    \item Giants: $>100~M_\oplus$
\end{itemize}
The occurrence rates are calculated as the number of planets in each category divided by the total number of systems (5000 in our simulations). NGPPS values are taken from \cite{Burn+2021}.

Figure~\ref{fig:statics} compares the occurrence rates across different stellar masses. We include additional simulations using a planetesimal radius of 300~m, matching NGPPS, for stellar masses of 0.1, 0.3, 0.5, and 1~$M_\odot$ (blue points).

When using the same planetesimal radius (compare the blue and black lines), our model yields occurrence rates for Earth-like, super-Earth, and Neptunian planets that closely match NGPPS across all stellar masses. In contrast, our nominal case with $R_{\rm plt} = 10$~km underestimates Neptunian occurrence around M dwarfs due to slower core growth. These results suggest that both models produce similar populations for planets that have not undergone runaway gas accretion.

However, our model predicts more sub-giants and giants than NGPPS, even with the same planetesimal radius. In fact, sub-giant occurrence in our model is comparable to or exceeds that in NGPPS for $M_* = 0.5$ and $1M_\odot$, even with $R_{\rm plt} = 10$~km. The relative abundance of sub-giants to giants is also higher in our model.

This difference arises from the treatment of orbital migration and runaway gas accretion. Our model includes disc gap opening via disc–planet interactions, which slows both type-II migration and gas accretion rates. 
Indeed, due to the relatively low mid-plane viscosity $\alpha_{\rm mid}$, even planets with a few Earth masses around the iceline can open gaps, transiting to a slower type-II regime. 
NGPPS, by contrast, uses classical type-II migration~\citep{Alexander+Armitage2009} and Bondi/Hill-like gas accretion~\citep{Dangelo+Lubow2008,Zhou+Lin2007}.
\rev{
Moreover, NGPPS adopts the mid-plane viscosity of $\alpha_{\rm mid} = 2\times 10^{-3}$, which is higher than $5\times 10^{-4}$ used in this study, which also leads to more efficient type-II migration in NGPPS model.
}

In our model, slower migration allows planets more time to accrete gas via Kelvin–Helmholtz contraction. Once runaway gas accretion begins, growth is limited by disc gas supply, and planets typically reach Jovian masses without significant orbital change~\citep{Tanaka+2020}. In NGPPS, planets often migrate to the disc inner edge before accreting substantial gas, while those that underwent runaway accretion grow rapidly to several Jovian masses.

In summary, our model produces similar distributions for non-giant planets (terrestrial and super-Earths) compared to NGPPS, but differs significantly in the formation frequency and final masses of giant planets. Note that the computation time for our population synthesis calculations is typically shorter by a factor of $\sim 10^5$ compared to that for NGPPS calculations.

\subsection{Revisits of planetary water abundance}

\begin{table}[t]
    \centering
    \caption{Fraction of planets in the habitable zone (HZ) around stars of different masses ($0.1M_\odot$, $0.3M_\odot$, $0.5M_\odot$). Only planets with masses of 0.3--3$M_\oplus$ are considered. Results for $X_{\rm H_2O,mix}=0.8$ are shown for both this study and KI22. See text for the definitions of planet types.
    }
    \label{tab:water_fraction}    
    \begin{tabular}{c|c|c|c} \hline
     Stellar Type & Planet Type & this study [\%] & KI22 [\%]\\ \hline 
     \multirow{3}{*}{0.1$M_\odot$} & Ocean-rich & 85.1 & 51.8 \\
                                   & Moderate   & 0.2  & 1.2  \\
                                   & Dry        & 14.7 & 47.0  \\ \hline   
     \multirow{3}{*}{0.3$M_\odot$} & Ocean-rich & 84.3 & 48.8 \\
                                   & Moderate   & 1.4  & 3.3  \\
                                   & Dry        & 14.3 & 47.9  \\ \hline 
     \multirow{3}{*}{0.5$M_\odot$} & Ocean-rich & 75.5 & 55.4 \\
                                   & Moderate   & 6.2  & 7.0  \\
                                   & Dry        & 18.3 & 37.6  \\ \hline
    \end{tabular}
\end{table}

Here we examine the effects of the model updates on the water content of habitable-zone (HZ) planets, on which KI22 focused.
Following KI22, we select planets with masses of 0.3–3~$M_\oplus$ located in the HZ, defined as 0.03--0.07~au, 0.1--0.2~au, and 0.2--0.4~au for $0.1M_\odot$, $0.3M_\odot$, and $0.5M_\odot$ stars, respectively~\citep{Kopparapu+2014}.  
Planets are categorised as Ocean-rich, Moderate, or Dry depending on their water mass fraction: Ocean-rich planets exceed 100 times the terrestrial ocean fraction ($=0.023$~wt.\%); Dry planets contain no water; and Moderate planets fall between these two limits.

Table~\ref{tab:water_fraction} compares the fractions of each type obtained with $X_{\rm H_2O,mix}=0.8$ in this study with those from KI22.  
Our updated model predicts a substantially larger fraction of Ocean-rich planets and a corresponding reduction in Dry planets compared with KI22. This change arises from the longer disc lifetimes assumed here, which allow more efficient inward migration of icy planets formed beyond the iceline. These planets displace in-situ rocky planets towards the inner regions, so that the HZ becomes dominated by water-rich planets originating beyond the iceline.  
\rev{
Note that the planetesimal composition depends sensitively on the disc temperature at the time of planetesimal formation, which is before the start time of our simulation, and is itself uncertain. Consequently, the fraction of icy planetesimals and the efficiency of inward delivery of water-rich material would depend on details of the early disc thermal evolution.
}

The fraction of Moderate planets is reduced relative to KI22, also owing to the ocean-rich planet migrations. 
Nonetheless, a few percent of planets around $0.3M_\odot$ and $0.5M_\odot$ stars still retain moderate water contents. As in KI22, this water is primarily supplied through envelope enrichment. The fraction of Moderate planets increases with stellar mass, again consistent with KI22. This trend reflects the fact that planets around more massive stars can grow larger before disc dispersal, allowing more envelope accretion, and that planets in the HZs around lower-mass stars experience stronger XUV irradiation.

In summary, our updated model alters the overall distribution of water abundances in the HZs of M dwarfs, strongly enhancing the prevalence of Ocean-rich planets. However, the essential result from KI22 remains robust: envelope enrichment provides a viable pathway for the formation of Earth-like planets with moderate water inventories. Our simulations still predict that several percent of Earth-mass planets in the HZs of mid- to early-M dwarfs possess water contents comparable to that of Earth.

\subsection{Caveats}
\label{sec:caveats}
Our planet population synthesis model introduces two key features: envelope enrichment with water vapour, which is assumed to occur via magma-gas chemical interactions, and a semi-analytical framework for planet-planet gravitational interactions.  
These are essential for interpreting exoplanet observations, including atmospheric compositions, and for enabling large-scale statistical simulations.  
However, both treatments involve several caveats.

For the enrichment process, we assume highly efficient water production and adopt a uniform water mass fraction $X_{\rm H_2O,mix}$ throughout the accreting envelope.
Chemical equilibrium calculations show that reduction of FeO by hydrogen can yield large water fractions, with H$_2$/H$_2$O mole ratio approaching unity ($X_{\rm H_2O,mix}\sim0.8$–0.9)~\citep{Ikoma+Genda2006,Kite+etal2020,Seo+etal2024}.
Recent high-pressure, high-temperature experiments further support this: substantial water production occurs under $\sim$50~GPa and $\gtrsim$4000~K conditions relevant to the magma-atmosphere boundary during formation~\citep{Miozzi+etal2025}, and even FeO-poor magmas can generate similarly high water fractions through reduction of silicates by metallic iron and hydrogen~\citep{Horn+etal2025}.
Together, these results indicate that substantial water enrichment is thermodynamically favoured over a wide range of conditions.

However, the actual envelope composition depends on the magma redox state~\citep[e.g.][]{Schaefer+etal2016,Kite+etal2020,Lichtenberg+etal2021,Schlichting+Young2022}, which reflects the planetesimal compositions.
Incorporating spatially varying solid compositions from condensation models~\citep[e.g.][]{Thiabaud+2014,Marboeuf+2014b,Marboeuf+2014a} would therefore provide more physically grounded estimates of $X_{\rm H_2O,mix}$ and improve the ability to link observed envelope properties to formation pathways.

%Furthermore, several neglected processes related to envelope enrichment would influence the resulting envelope mass and composition.
\rev{
Another major assumption is that the H–He layer accreted in Phase~II does not mix with the underlying vapour-mixed layer. This represents a limiting case in which a strong compositional gradient suppresses large-scale convection. This assumption must be verified: From the thermodynamic point of view, H$_2$ and H$_2$O are miscible at the interface with sufficiently high pressures and temperatures~\citep{Soubiran+Militzer2015,Bergermann+etal2024,Gupta+etal2025}. Efficient mixing would increase the envelope mass in enriched cases, implying that our treatment likely underestimates the impact of water enrichment on envelope growth \citep[see][]{Hori+Ikoma2011,Kimura+Ikoma2020}.
}

For planets forming beyond the iceline, enrichment can also proceed via sublimation of accreting ices, which has a large impact on the envelope accumulation process~\citep[e.g.][]{Venturini+2016,Ormel+2021}.
In addition, we currently ignore additional processes acting on enriched envelopes, such as water dissolution into magma~\citep[e.g.][]{Papale1997,Kite+etal2020,Lichtenberg+etal2021}.
%and mixing between H-He and vapour-mixed layers, which is likely efficient under the high-pressure, high-temperature conditions~\citep[e.g.][]{Soubiran+Militzer2015,Bergermann+etal2024,Gupta+etal2025}.
Future work will incorporate more detailed treatments of envelope enrichment, including magma-envelope redox reactions, ice evaporation, and water dissolution, to better assess their impact on planetary properties.

Regarding dynamical evolution, our semi-analytical model is less accurate for distant planets (e.g., beyond a few au around solar-mass stars) and systems with giant planets, where gravitational scattering and ejection dominate~\citep{Kimura+2025b}.  
In such cases, the model tends to overestimate eccentricity excitation, leading to inflated rates of ejection, collision, and semi-major axis changes.  
While this has limited impact on close-in planets, which are the current focus of most exoplanet surveys, the number of detections at wider orbits is expected to grow with missions such as PLATO and Roman.  
Improving predictions for distant planets is therefore increasingly important, and extending our model to better treat such systems is necessary to enhance the applicability of our model.

\section{Conclusions}
\label{sec:conclusion}

We have updated our planetary population synthesis model to incorporate several key improvements over our previous framework (KI22), aiming to more accurately capture the physical processes that govern planet formation and evolution. The major updates include a revised treatment of disc evolution model based on stellar X-ray photoevaporation, yielding a broader and more realistic distribution of disc lifetimes, and a semi-analytical model for post-disc dynamical interactions, enabling efficient and statistically consistent modelling of giant impacts and gravitational scattering.

These updates have a significant impact on the resulting planetary populations. Compared to KI22, our model predicts:
\begin{itemize}
    \item Reduced abundance of low-mass planets (e.g. Mars-like bodies) in close-in orbits due to more frequent collisional growth after disc dispersal.
    \item A broader mass distribution of giant planets, including a substantial population of super-Jovian planets formed in long-lived discs.
    \item More efficient inward migration and stellar engulfment, leading to an increased number of short-period planets.
\end{itemize}

We also investigated the effects of envelope enrichment and stellar mass on the final planet distributions. Water enrichment in the envelope significantly increases the retention of atmospheres in Earth-like to super-Earth mass planets, resulting in distinct clusters in the radius distribution. It also enhances the formation frequency of giant planets.

Stellar mass strongly influences both the typical planet mass and the envelope retention. Around low-mass stars ($\lesssim 0.3M_\odot$), planets tend to be smaller and more vulnerable to photoevaporation, while around more massive stars, planets can grow larger and retain more diverse envelope compositions. In particular, we find that 1–5\% of Earth-mass planets in the habitable zones of mid- to early-M dwarfs are predicted to have water inventories ranging from 0.01 to 100 times that of Earth, consistent with our previous findings in KI22.

We further compared our results with those from the New Generation Planet Population Synthesis (NGPPS). While both models yield similar distributions for non-giant planets, our model predicts higher occurrence rates of sub-giant and giant planets. This difference arises from the slower orbital migration and gas accretion rates in our model, which allow more planets to enter the runaway gas accretion phase and grow to Jovian masses without migrating to the disc inner edge.

In summary, the updated model offers a more physically consistent framework for simulating planetary system formation and highlights the importance of envelope composition during the formation stage in shaping final planetary architectures. 
Moreover, because of its high computational efficiency, the model enables population synthesis across wide parameter ranges and the generation of a large number of planet samples within feasible computational time. 
This capability allows for validating whether current formation theories can reproduce the observed exoplanet population and enables to quantify how individual parameters influence planetary distributions. Such quantitative comparisons with observations will be presented in our subsequent paper.

\begin{acknowledgements}
    This work was supported by JSPS KAKENHI Grant No. JP22K21344 and Daiichi-Sankyo "Habataku" Support Program for the Next Generation of Researchers.
    Part of the numerical computations were carried out on the general-purpose PC cluster at the Center for Computational Astrophysics, National Astronomical Observatory of Japan, and 
    on the Hábrók high performance computing cluster at the University of Groningen.
\end{acknowledgements}

\bibliographystyle{aa} % style aa.bst
\bibliography{refer} % your references Yourfile.bib

\appendix

\section{Asymptotic Mass}
\label{sec:asymptotic_mass}

Here, we derive an analytical expression for the planetary mass that can be acquired via in-situ planetesimal accretion. This is used to determine the initial locations of planetary embryos in our simulations (\S~\ref{sec:init_cond_planet}).

In the inner disc regions (typically inside the iceline), planetary growth is limited by the local isolation mass. In contrast, in the outer regions, the integration time ($\sim$~Gyr) is insufficient to reach isolation. In such cases, growth before disc dispersal can be neglected, and only post-dispersal growth needs to be considered. According to \S~\ref{sec:solid_core_growth}, the growth rate is given by
\begin{equation}
    \dv{M_\text{core}}{t} = \Omega_\text{K} \bar{\Sigma}_\text{s} R_\text{H}^2 p_\text{col}.
\end{equation}
The collision probability $p_{\rm col}$ from \cite{Inaba+2001} depends on planetesimal eccentricities. After disc dispersal, eccentricities increase to approximately $e_{\rm esc}$, so the high-eccentricity regime applies:
\begin{align}
    p_{\rm col} = p_\text{col,high} = \frac{1}{2\pi} \qty(\frac{R_\text{cap}}{R_\text{H}})^2
    \qty(
    \mathcal{F}(i_\text{plt}/e_\text{plt}) + \frac{6R_\text{H}\mathcal{G}(i_\text{plt}/e_\text{plt})}{R_\text{cap}\tilde{e}_\text{plt}^2}
    ),
\end{align}
\rev{
where $\tilde{e}_{\rm plt} = e_{\rm plt}/h$ is the reduced eccentricity, with $h = (M_{\rm core}/3M_*)^{1/3}$.
}
Assuming $i_{\rm plt} = e_{\rm plt}/2$, we adopt $\mathcal{F} = 17.3$ and $\mathcal{G} = 38.2$, yielding
\begin{equation}
    p_{\rm col} = \frac{1}{2\pi} \qty(\frac{R_{\rm core}}{R_{\rm H}})^2
    \qty( 17.3 + \frac{229.2R_{\rm H}}{R_{\rm core} \tilde{e}_{\rm plt}^2}).
    \label{eq:pcol}
\end{equation}
Here, $R_{\rm cap} = R_{\rm core}$ is assumed.
Substituting the escape eccentricity,
\begin{equation}
    e_{\rm plt} = e_{\rm esc} = \frac{\sqrt{2GM_{\rm core}/R_{\rm core}}}{\sqrt{GM_*/a_{\rm p}}} = \sqrt{6h^2\frac{R_{\rm H}}{R_{\rm core}}},
    \label{eq:eplt_esc}
\end{equation}
%with $h = (M_{\rm core}/3M_*)^{1/3}$, 
we obtain a simplified form:
\begin{equation}
    p_{\rm col} \simeq \frac{55.5}{2\pi}\qty(\frac{R_{\rm core}}{R_{\rm H}})^2.
\end{equation}
Then, the growth rate becomes
\begin{equation}
    \dot{M}_{\rm core} = \frac{55.5}{2\pi} \Omega_{\rm K}\Sigma_{\rm s}R_{\rm core}^2
    =  \frac{55.5}{2\pi} \Omega_{\rm K}\Sigma_{\rm s} \qty(\frac{3}{4\pi \rho_{\rm p}})^{2/3}M_{\rm core}^{2/3}
    =: 3A\Sigma_{\rm s} M_{\rm core}^{2/3},
    \label{eq:dM_dt_appx}
\end{equation}
where $\rho_{\rm p}$ is the bulk density. This leads to
\begin{equation}
    \dv{M_{\rm core}^{1/3}}{t} = A\Sigma_{\rm s}.
\end{equation}

To account for the depletion of solids due to planetary growth, we express
\begin{equation}
    \Sigma_{\rm s}= \Sigma_{\rm s0} -\frac{M_{\rm core}}{2\pi a \Delta a_{\rm FZ}} 
    = \Sigma_{\rm s0}- \frac{3^{1/3}M_*^{1/3}}{2^{4/3}\pi b_{\rm FZ}a^2}M_{\rm core}^{2/3}
    =: \Sigma_{\rm s0}- C M_{\rm core}^{2/3},
\end{equation}
and the differential equation becomes
\begin{equation}
    \dv{M_{\rm core}^{1/3}}{t} = A\Sigma_{\rm s0} - ACM_{\rm core}^{2/3}.
    \label{eq:analytic_dMc}
\end{equation}
Integrating this with the initial condition $M_{\rm core} = 0$ at $t = 0$, we obtain
\begin{equation}
    M_{\rm core}^{1/3} = \sqrt{\frac{\Sigma_{s0}}{C}}\tanh(A\sqrt{C\Sigma_{\rm s0}}t),
\end{equation}
which asymptotically approaches the isolation mass as $t \to \infty$.

Thus, the asymptotic mass due to in-situ planetesimal accretion is
\begin{equation}
    M_{\rm asym} = \qty(\frac{\Sigma_{s0}}{C})^{3/2}
    \tanh^3(A\sqrt{C\Sigma_{\rm s0}}t_{\rm max}).
    \label{eq:M_asym}
\end{equation}

In addition, growth of distant planets may be limited by planetesimal ejection, which becomes significant when $e_{\rm esc} > 1$, corresponding to
\begin{equation}
    M_{\rm p} > M_{\rm esc} := \sqrt{\frac{3M_*^3}{32\pi \rho_{\rm p} a_{\rm p}^3}},
    \label{eq:M_esc}
\end{equation}
assuming a bulk density $\rho_{\rm p} = 3~{\rm g/cm^3}$.

The maximum mass attainable via in-situ planetesimal accretion is then
\begin{equation}
    M_{\rm max} = \min\qty( M_{\rm asym}, M_{\rm esc}),
    \label{eq:M_max}
\end{equation}
which is used to determine the initial placement of planetary embryos.

\rev{
\section{Pre-computed Structure Grids}
\label{sec:grids}
}

\rev{
To accelerate the thermal evolution calculations, we pre-compute
multidimensional grids of internal structure models and interpolate within them during population synthesis.
Separate grids are constructed for
(i) thin envelopes (\S~\ref{sec:evolution_thin_envelope}),
(ii) thick H–He envelopes (\S~\ref{sec:evolution_thick_envelope}), and
(iii) pure water layers (\S~\ref{sec:evolution_water_layer}).
Below we summarise their parameter coverage and resolution.
}

\rev{
\subsection{Thin-envelope grid}
\label{sec:grid_thin}
}

\rev{
For planets that have not undergone runaway gas accretion,
we tabulate $(R_{\rm p}, T_{\rm surf}, E_{\rm env})$
as functions of six input parameters:
\[
(X_{\rm H_2O}, \log T_{\rm eq}, \log M_{\rm core},
\log F_{\rm mix}, \log F_{\rm HHe}, \log L),
\]
where $F_{\rm mix}=M_{\rm mix}/M_{\rm c}$
and $F_{\rm HHe}=M_{\rm HHe}/M_{\rm c}$.
All quantities except $X_{\rm H_2O}$ are sampled in logarithmic space.
The adopted ranges and resolutions are:
\begin{itemize}
\item $X_{\rm H_2O}$: 0.0–0.9 (step 0.1),
\item $\log T_{\rm eq}$ [K]: 2.0–3.4 (step 0.2),
\item $\log M_{\rm core}$ [$M_\oplus$]: $-1.0$–2.0 (step 0.25),
\item $\log F_{\rm mix}$: $-5$ – $-0.5$ (step 0.5),
\item $\log F_{\rm HHe}$: $-10$ and $-5$–0.0 (step 0.5),
\item $\log L$ [erg s$^{-1}$]: 19–26 (step 0.2).
\end{itemize}
This corresponds to $\sim 4.5\times10^5$ structure calculations
per $X_{\rm H_2O}$ value and $\sim 4.5\times10^6$ models in total.
During population synthesis,
radii and surface temperatures are obtained via linear interpolation
in this six-dimensional space, without extrapolation.
}

\begin{figure}
    \centering
    \includegraphics[width=\linewidth]{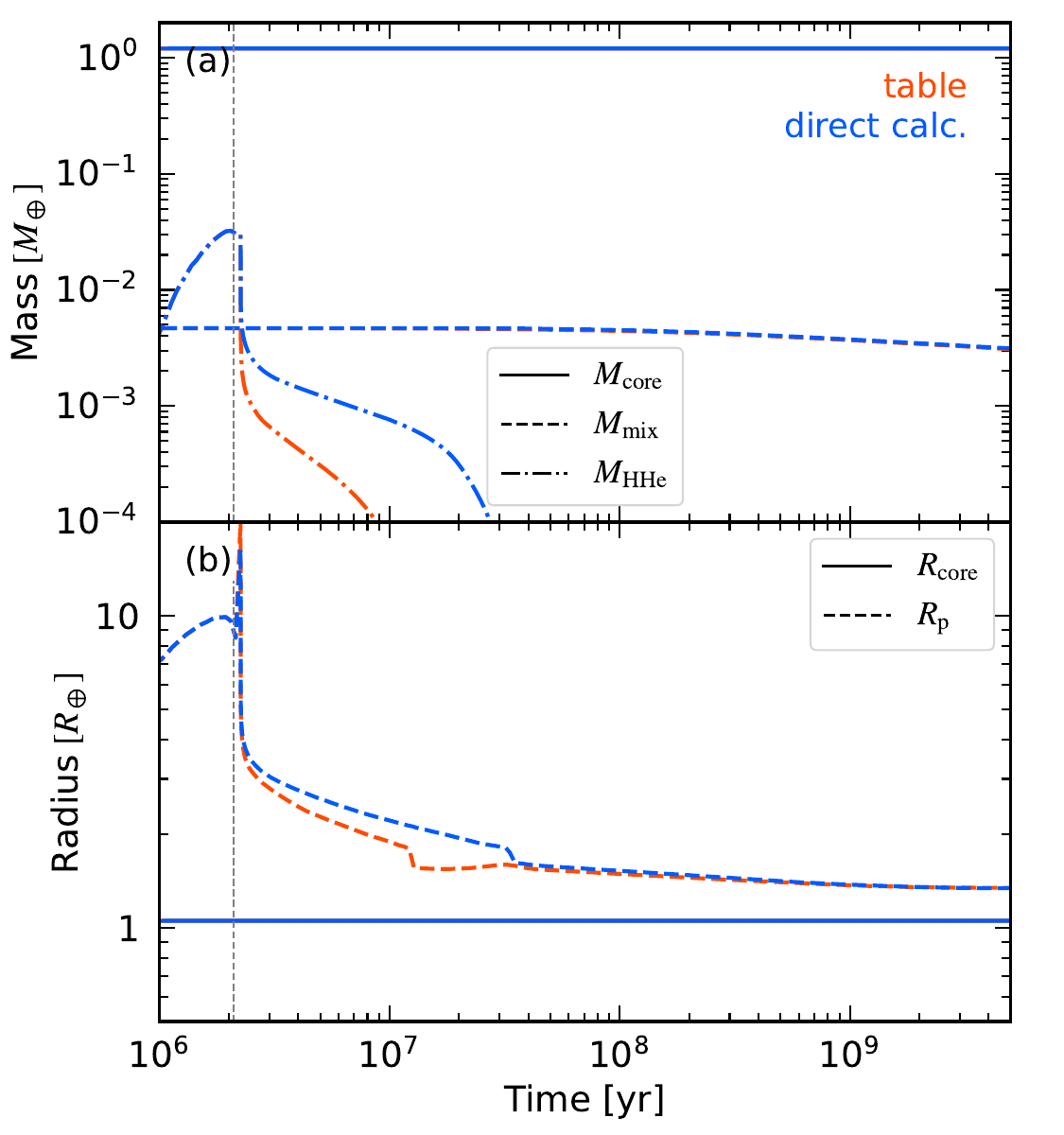}
    \caption{
    \rev{
    Evolution of the mass and radius of a planet with the same parameters and initial conditions as in Fig.~\ref{fig:one_planet}. Here the case with $X_{\rm H_2O,mix}=0.5$ is shown.
    The red and blue lines show the results with using the pre-calculated table and with directly calculating the post-disc thermal evolution of the envelope, respectively.
    The vertical dashed line indicates the disc dispersal time.
    }
    }
    \label{fig:evol_table_vs_direct}
\end{figure}

\rev{
To assess the interpolation accuracy, we computed the post-disc thermal evolution and loss of the envelope for a planet shown in Fig.~\ref{fig:one_planet}, with the case of $X_{\rm H_2O,mix}=0.5$, by directly integrating the internal structures.
Figure~\ref{fig:evol_table_vs_direct} compares the results with the one presented in Fig.~\ref{fig:one_planet}.
We found that the dissipation timescale of H-He envelope becomes longer when the envelope thermal evolution is directly calculated, although the radius is quite similar to our nominal case that uses the table.
This is because the escape rate of the inflated envelope at the beginning of the post-disc phase, given by \cite{Kubyshkina+2018a}, is quite highly sensitive to the planet radius. Thus, the small change in the radius in such early stage largely affects the dissipation timescale.
However, the impact on the long-term sustainability of the envelope is small, and the resultant planetary radius at the final state is sufficiently similar to each other,
demonstrating that the adopted resolution is sufficient for statistical population studies.
}

\rev{
\subsection{Thick-envelope grid}
\label{sec:grid_thick}
For planets that have undergone runaway gas accretion,
we tabulate $(R_{\rm p}, T_{\rm surf})$
as functions of
$(T_{\rm eq}, \log M_{\rm c}, \log M_{\rm env}, \log t)$.
The grid ranges are:
\begin{itemize}
\item $T_{\rm eq}$ [K]: 100–2000 (step 100),
\item $\log M_{\rm c}$ [$M_\oplus$]: 0.0–2.0 (step 0.2),
\item $\log M_{\rm env}$ [$M_\oplus$]: $-1$–4.0 (step 0.2),
\item $\log t$ [yr]: 6–10 (step 0.25).
\end{itemize}
This yields $\sim 9.7\times10^4$ grid models.
Given the smoother thermodynamic structure of fully convective envelopes,
this resolution is sufficient to capture the long-term radius evolution.
}

\rev{
\subsection{Water-layer grid}
\label{sec:grid_water}
For planets that have lost their H–He envelopes,
we compute a separate grid for pure water layers.
We tabulate $(R_{\rm p}, T_{\rm surf})$
as functions of
$(\log T_{\rm eq}, \log M_{\rm tot}, M_{\rm ice}/M_{\rm tot}, \log t)$,
where $M_{\rm tot}=M_{\rm rock}+M_{\rm ice}$.
The parameter coverage is:
\begin{itemize}
\item $\log T_{\rm eq}$ [K]: 2.0–3.4 (step 0.2),
\item $\log M_{\rm tot}$ [$M_\oplus$]: $-1.0$–2.0 (step 0.25),
\item $M_{\rm ice}/M_{\rm tot}$: 0.05–0.5 (step 0.05),
\item $\log t$ [yr]: 6–10 (step 0.25).
\end{itemize}
This corresponds to $\sim 1.8\times10^4$ models.
Linear interpolation within this grid is used
to determine planetary radii after complete envelope loss.
}

\end{document}